\documentclass[sigconf]{acmart}

\usepackage{multirow}
\usepackage{color}
\usepackage{colortbl}
\usepackage{stfloats}
\usepackage{subcaption}
\usepackage{enumitem}
\usepackage{graphicx}
\usepackage{subcaption}
\usepackage{tabularx}
\usepackage{longfbox}
\usepackage{soul}

\makeatletter
\newdimen\@tempdimd
\makeatother

\newfboxstyle{boxparam}{padding=1.5pt, padding-left=2pt, padding-right=2pt, height=7.5pt, background-color=lightgray, border-radius=2pt, border-right-width=1pt, border-bottom-width=1pt, border-color=gray}

\newcommand{\rectwrapsmall}[2]{\lfbox[boxparam, border-radius=0pt, padding-left=2pt, padding-right=2pt, height=5.5pt, border-width=0pt, background-color=#1]{\sffamily{\textcolor{white}{#2}}}}

\newcommand{\blackrectsmall}[1]{\rectwrapsmall{darkgray}{#1}}

\definecolor{quotebackground}{HTML}{EFEFEF}
\definecolor{tableheader}{HTML}{EFEFEF}
\definecolor{tablegrayline}{HTML}{e0e0e0}

\newcommand{\sysname}{\textsc{ExploreSelf}}

\newcommand{\exampleusername}{Jane}

\newcommand{\eg}{\textit{e.g.}}
\newcommand{\ie}{\textit{i.e.}}
\newcommand{\cf}{\textit{c.f.}}
\newcommand{\etal}{\textit{et al.}}
\newcommand{\pval}[1]{$p = #1$}
\newcommand{\reportstats}[5]{$SD=#1$, $min=#2$ [#3], $max=#4$ [#5]}

\newboolean{clean}
\setboolean{clean}{true} 

\newcommand{\revised}[1]{\ifthenelse{\boolean{clean}}{#1}{\textcolor{blue}{#1}}}

\newboolean{cameraclean}
\setboolean{cameraclean}{true} 

\newcommand{\cameraready}[1]{\ifthenelse{\boolean{cameraclean}}{#1}{\textcolor{blue}{#1}}}

\newenvironment{revisedenv}{%
    \begingroup
    \ifthenelse{\boolean{clean}}{}{%
        \color{blue}%
    }%
}{%
    \endgroup
}

\newcommand{\deleted}[1]{%
  \ifthenelse{\boolean{clean}}{}{%
    \textcolor{cyan}{\st{#1}}%
  }%
}

\newcommand{\deletedsubsection}[1]{%
  \ifthenelse{\boolean{clean}}{}{%
    \subsection{\textcolor{cyan}{[Deleted] #1}}%
  }%
}

\newcommand{\circledigit}[1]{\textbf{\normalsize{\textsf{\textcircled{\footnotesize{#1}}}}}}

\newcommand{\ipstart}[1]{\vspace{1mm} \noindent{\textbf{\textit{#1.}}}}

\newcommand{\npstart}[1]{\vspace{1mm} \noindent{{#1}}}

\newcommand{\sourcecode}{\urlstyle{tt}\url{https://naver-ai.github.io/exploreself}}

\copyrightyear{2025}
\acmYear{2025}
\setcopyright{rightsretained}
\acmConference[CHI '25]{CHI Conference on Human Factors in Computing Systems}{April 26-May 1, 2025}{Yokohama, Japan}
\acmBooktitle{CHI Conference on Human Factors in Computing Systems (CHI '25), April 26-May 1, 2025, Yokohama, Japan}
\acmDOI{10.1145/3706598.3713883}
\acmISBN{979-8-4007-1394-1/25/04}

\begin{document}

\title[\sysname{}: Fostering User-driven Exploration and Reflection on Personal~Challenges with Adaptive Guidance by LLMs]{\sysname{}: Fostering User-driven Exploration and Reflection on Personal~Challenges with Adaptive Guidance by Large~Language Models}


\author{Inhwa Song}
\authornote{Inhwa Song conducted this work as a research intern at NAVER AI Lab.}
\orcid{0009-0000-2325-663X}
\affiliation{%
  \institution{KAIST}
  \country{Republic of Korea}
}
\email{inhwa.song@kaist.ac.kr}

\author{SoHyun Park}
\orcid{0000-0001-8703-0584}
\affiliation{%
  \institution{NAVER Cloud}
  \country{Republic of Korea}}
\email{sohyun@snu.ac.kr}

\author{Sachin R. Pendse}
\orcid{0000-0001-6925-3258}
\affiliation{
    \institution{Northwestern University}
    \country{Evanston, IL, USA}
}
\email{sachin.r.pendse@northwestern.edu}

\author{Jessica Lee Schleider}
\orcid{0000-0003-2426-1953}
\affiliation{
    \institution{Northwestern University}
    \country{Evanston, IL, USA}
}
\email{jessica.schleider@northwestern.edu}

\author{Munmun De Choudhury}
\orcid{0000-0002-8939-264X}
\affiliation{
    \institution{Georgia Institute of Technology}
    \country{Atlanta, GA, USA}
}
\email{munmund@gatech.edu}

\author{Young-Ho Kim}
\orcid{0000-0002-2681-2774}
\affiliation{%
  \institution{NAVER AI Lab}
  \country{Republic of Korea}
}
\email{yghokim@younghokim.net}

\begin{abstract}
Expressing stressful experiences in words is proven to improve mental and physical health, but individuals often disengage with writing interventions as they struggle to organize their thoughts and emotions. Reflective prompts have been used to provide direction, and large language models (LLMs) have demonstrated the potential to provide tailored guidance. However, current systems often limit users' flexibility to direct their reflections. We thus present \sysname{}, an LLM-driven application designed to empower users to control their reflective journey, providing adaptive support through dynamically generated questions. Through an exploratory study with 19 participants, we examine how participants explore and reflect on personal challenges using \sysname{}. Our findings demonstrate that participants valued the flexible navigation of adaptive guidance to control their reflective journey, leading to deeper engagement and insight. Building on our findings, we discuss the implications of designing LLM-driven tools that facilitate user-driven and effective reflection of personal challenges.
\end{abstract}



\begin{teaserfigure}
    \centering
    \includegraphics[width=\textwidth]{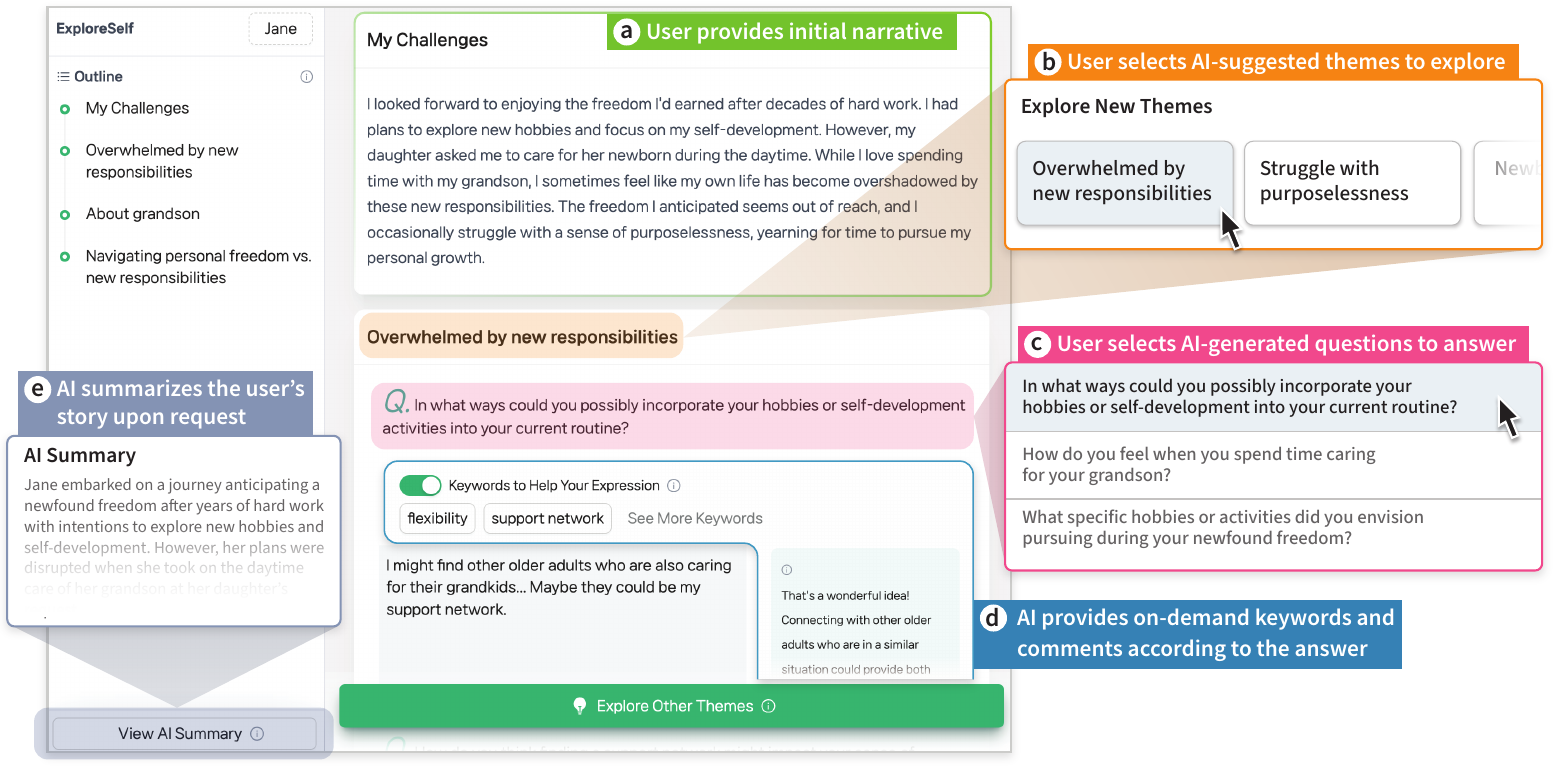}
    \caption[
    The figure illustrates how ExploreSelf supports user-driven exploration and reflection on personal challenges. It shows the process beginning with the user's initial narrative (denoted as a), which serves as the foundation for further exploration. From this starting point, the user can choose to explore different perspectives by following AI-suggested or custom themes (b). For each theme, the system generates specific questions (c) that guide the user's reflection. To support this process, ExploreSelf provides adaptive guidance in the form of AI-generated keywords and comments (d) that users can reference as they write their responses. At any point, the user can request an AI-generated summary of the current exploration (e).
    ]{\sysname{} supports user-driven exploration and reflection on their personal challenges. \revised{Based on the initial narrative~\circledigit{a} written by the user, they can explore diverse perspectives by following up the AI-suggested or custom themes~\circledigit{b}. Regarding each theme, the user can reflect on detailed aspects through questions generated by an AI~\circledigit{c}. As adaptive guidance, the AI provides keywords and comments on the fly that the user can refer to while writing an answer~\circledigit{d}. At any time, the user can request an AI-generated summary of the current exploration~\circledigit{e}.} 
    }
    \label{fig:interface:teaser}
\end{teaserfigure}

\maketitle

\section{Introduction}
Articulating stressful or emotional experiences in words has been found to improve both physical and mental health by enabling individuals to process and express their thoughts and emotions~\cite{baikie2005emotional, pennebaker1993putting, pennebaker1997writing}. It has been a widely circulated and popular means for individuals to self-reflect and make sense of themselves~\cite{hubbs2005paper, evans2011authoring}, by simply sitting down to write the innermost thoughts and feelings without confines of grammar, punctuation, or any other formalities of written composition. However, writing about negative experiences and thoughts can be challenging as it often requires individuals to confront distressing thoughts and emotions, making it tempting for them to disengage from the process~\cite{smyth2002does, pennebaker1997writing}. Hence, there have been a number of adjustments to the activity to help with this process. For example, tools like self-help workbooks and prompted journaling have been developed to offer structured prompts that guide users in exploring their thoughts and emotions~\cite{moussa2015reflective, regan2005promoting}. The HCI community has also introduced interactive methods, such as using photos as journaling prompts~\cite{karaturhan2022combining} or employing a conversational assistant with sequences of probing questions~\cite{park2021wrote}. These approaches mostly involved pre-defined prompts for writers to use, which has proven effective to spark thinking and dive into the process~\cite{davis2003prompting, kauffman2008prompting}. It has also been noted that these prompts are inflexible and detached from the user context~\cite{blunk2017prompting}.

The generative capabilities of recent large language models (LLMs) have accelerated the advancement of intelligent writing support tools~\cite{lee2024designspacein2writing, lee2022coauthor}, which commonly perform natural language generation for directly modifying the content (\eg{},~\cite{kim2024diarymate, lee2022coauthor}) or providing advice for improving the content (\eg{}, \cite{evmenova2024improving}). However, research on LLM-driven writing tools largely focuses on creative writing, and only a handful of works have recently begun to study intelligent support for writing about the self (\eg{}, DiaryMate~\cite{kim2024diarymate}). Further, a growing body of research has investigated the LLMs' potential for mental health support in the form of conventional chat interfaces~(\eg{},~\cite{kim2024mindfuldiary, seo2024chacha, Kumar2024selfreflection, song2024typing}). Although LLM-driven, open-ended conversations are intuitively analogous to therapists' counseling sessions, an agent who holds the therapist's persona tends to lead the conversation and sometimes can result in unintentional reinforcement of negative narratives~\cite{song2024typing, PiChatbot, kim2024mindfuldiary}. Meanwhile, research suggests that when individuals are able to make choices and exert control over their experiences, they engage more meaningfully in the process and develop a stronger sense of ownership and responsibility for their growth~\cite{deci2008self}.

In this work, we investigate ways to empower individuals to take greater control of the reflective process of their personal challenges while simultaneously leveraging the personalized and adaptive guidance offered by LLMs.
\cameraready{Unlike traditional mental health services that focus on professional diagnosis and intervention, our goal was to design a system that enables individuals to explore their challenges at their own pace, without clinical oversight.}
To this aim, we designed and developed \sysname{}~\revised{(\autoref{fig:interface:teaser})}, an LLM-infused interactive system for exploring and reflecting on personal challenges. \sysname{} allows users to freely expand related themes from the initial narrative and drill down into probing questions to reflect on aspects of themes in depth. To facilitate exploration, the system provides on-demand guidance, such as prompting keywords and comments tailored to their writing. The design of \sysname{} was informed by formative interviews with nine mental health professionals, where their challenges and approaches to managing the counseling sessions inspired our interface design and the generative pipelines.

To understand how people interact and engage with \sysname{} in exploring their personal challenges, we conducted an online lab study with 19 individuals, where they explored their personal challenges using \sysname{} for 30 to 60 minutes. 
Through user-directed exploration with diverse forms of adaptive guidance by LLMs, \sysname{} enabled individuals to engage in meaningful self-reflection while maintaining control over the direction of their reflection. Participants actively leveraged the navigation components (\ie{}, themes, questions, and summary) and the adaptive guidance (\ie{}, keywords, and comments) in individualized patterns. Participants' perceived agency was significantly increased after the session. Participants reported that the system’s adaptive features, such as the ability to select themes, choose from multiple Socratic questions, and toggle keywords and comments, contributed to their sense of control and autonomy. Participants decided which aspects to focus on or revisit later in a way that balances their cognitive and emotional effort with the expected significance of the engagement regarding their personal concerns. Drawing on the findings, we discuss how the components of \sysname{} collectively fostered a dynamic and personalized reflective process, helping people feel more empowered in navigating their personal challenges.

\npstart{The key contributions of this work are threefold:}
\begin{enumerate}[leftmargin=*, itemsep=4pt, topsep=0pt]
\item The design and implementation of \sysname{}, a novel LLM-driven web application that supports guided writing about breakdown topics regarding a personal issue, encouraging sensemaking and refreshment of thoughts on it. \sysname{}'s design was informed by formative interviews ($N=9$) with clinical psychologists and professional licensed counselors. The source code of \sysname{} is publicly available at \sourcecode{}.
\item An empirical study using \sysname{}, with 19 individuals with varied severity and types of personal issues. From the quantitative and qualitative analysis, we provide an understanding of how people elaborate thoughts and make sense of their own issues through interactive reflection aided by adaptive guidance. We uncover how participants balanced exploration and reflection, engaging with AI-driven themes, questions, keywords, and comments, and offer insights into their perspectives on these forms of guidance.

\item Implications for designing LLM-guided tools that facilitate user-driven and effective reflection of personal challenges, \revised{such as balancing user autonomy with guided support, fostering long-term practice, and implementing the nuanced role and stance of AI to accommodate a diverse range of individuals.} 
\end{enumerate}

\section{Related Work}
In this section, we cover the related work in the areas of (1) writing for well-being and (2) technology-mediated writing for well-being.

\subsection{Writing for Well-being}
A substantial body of research has demonstrated that the act of writing as an instrument for exploring personal thoughts and emotions can promote self-healing and personal growth, significantly enhancing both physical and emotional well-being~\cite{baikie2005emotional, pennebaker1993putting, pennebaker1997writing}. It has been widely studied across diverse contexts, including expressive writing from social psychology~\cite{ruini2022writing, pennebaker1997writing}, integration into specific psychotherapies as a treatment (\eg{}, guided autobiography, using diary in CBT), positive interventions to promote psychological well-being (\eg{}, writing about gratitude and forgiveness), and through personal practices~\cite{ruini2022writing}. One of the most widely researched methods is expressive writing, which involves writing about one’s deepest thoughts and feelings regarding stressful experiences~\cite{pennebaker1997writing}. Across these various writing for well-being practices, the common focus lies in translating personal experiences into written language, allowing individuals to process and organize their thoughts and feelings in a meaningful way~\cite{ruini2022writing, pennebaker1993putting,pennebaker1997writing,pennebaker1988disclosure}.

Through this process, individuals often derive meaning from stressful events and foster a sense of hope as they work through their emotions and gain new perspectives~\cite{harvey1990social}. This activity has been shown to yield significant benefits for individuals across various demographics, including different age groups~\cite{koschwanez2013expressive, travagin2015effective}, gender~\cite{manier2005benefits}, cross-cultures~\cite{kim2005effects, dominguez1995roles}, and health conditions~\cite{lu2012pilot, riddle2016does}. Studies have also found that writing can be beneficial for well-being when people engage in writing outside of controlled laboratory settings, such as at home or online~\cite{lepore2020expressive}. While several theories attempt to explain the connection between expressive writing and health, no single mechanism has emerged as dominant, likely due to the varied experiences and expectations of participants~\cite{langens2007effects}. However, studies from psychology have suggested that writing allows individuals to release suppressed emotions~\cite{lepore2002writing}, gain deeper insights into their experiences~\cite{pennebaker1993putting}, and develop a sense of control over their thoughts and emotions through self-reflection and emotion regulation~\cite{lepore2002writing}. In this way, writing can be used intentionally and purposefully to foster self-discovery, emotional growth, and personal insight, enabling individuals to gain a deeper understanding of themselves and their experiences~\cite{thompson2004journal,progoff1977journal,adams2009journal}. 

Research from psychology has indicated that writing about stressful events demands a significant level of perceived control, as it helps individuals organize, connect, and contextualize their thoughts and emotions~\cite{andersson2008expecting, frazier2011perceived}. This sense of control is not only essential for initiating or sustaining a writing practice but is also closely tied to physical and psychological well-being, making it a core component of the writing process itself~\cite{andersson2008expecting}. To support individuals in maintaining a sense of control, structured guidance has been studied as an effective tool to reduce cognitive load and alleviate the overwhelm that often accompanies reflection. Prompts and reflection frameworks are amongst the popularly studied tools for offering that sense of control, offering a sense of direction or goal to writers. In this way, individuals can navigate their thoughts and emotions in a more organized way, providing manageable steps to follow rather than leaving them to explore themselves entirely on their own. This approach supports sustained engagement by making the process more approachable and less overwhelming~\cite{cleary2004self, bandura1991social}. Previous research has explored various methods for designing such structure, including gratitude-focused writing ~\cite{fekete2022brief}, writing in a narrative structure~\cite{danoff2010does}, using sentence stems (\eg{}, ``\textit{The thing I am most worried about is...}''), journaling with photographs ~\cite{nave2016exploring}, and mind maps~\cite{danoff2010does, thompson2004journal}. While these approaches offer helpful starting points, they are often provided in a fixed format, usually requiring the individual to stay on a structured course throughout, therefore they cannot respond to the writer's thoughts and feelings that evolve in writing. This limits their ability to provide adaptive feedback or tailored suggestions to the individual's unique emotional states and goals in writing. Recognizing this gap, recent works have explored more interactive methods that provide personalized prompts, real-time feedback, and adaptive support~\cite{miller2014interactive, wang2018mirroru, jeong2016improving}. 

\subsection{Technology-Mediated Writing for Well-being}

With the soaring number of digital writing tools, individuals increasingly engage in writing for well-being using these tools, such as on online health platforms and digital journaling tools, to motivate themselves to engage in writing for well-being. ~\cite{ma2021detecting, smyth2008exploring}. These spontaneous, technology-mediated practices provide a space for self-expression and reflection, yet users often face challenges in maintaining consistent engagement and fostering deeper introspection. In response, the HCI community has explored methods to support individuals in their reflective writing journeys, focusing on strategies to enhance engagement by introducing the social presence of chatbots~\cite{park2021wrote} and providing contextual data to aid reflection~\cite{hodges2006sensecam}. These approaches have aimed to create structured yet flexible systems that guide users in navigating their thoughts and emotions more effectively.

The generative capabilities of LLMs (\eg{},  GPT~\cite{Brown2020FewShotLearners}, Gemini~\cite{team2023gemini}, Claude~\cite{claude3model}, HyperCLOVA~\cite{Kim2021HyperCLOVA}) have motivated the development of flexible and adaptive support for writing for well-being (\eg{},\cite{kim2024diarymate, kim2024mindfuldiary, sharma2024facilitating, Nepal2024journalingLLMSensing, yang2024psychotherapy}), offering personalized interactions that acknowledge and are responsive to what users write. For example, Kim~\etal{} explored the potential of interactive dialogue with LLMs as an alternative format for reflective journaling for psychiatric patients, where LLM chatbots provided adaptive reflective questions~\cite{kim2024mindfuldiary}. Their findings demonstrated that these adaptive prompts benefited individuals by helping them explore and reflect on their past experiences and emotions. However, 
since interactions with LLMs are naturally and intuitively conversational in nature, many LLM-driven approaches let users ``chat'' with an AI agent rather than write~\cite{PiChatbot, Replika, song2024typing, kim2024mindfuldiary}. Conventional turn-taking structure in chat inherently confines users to respond to the agent's last message, and such linearly progressing dialogues limit the flexibility to pivot subjects or grasp the landscape of the conversation. This makes the user's thought process heavily influenced by the performance and behavioral design of the AI agent, introducing a risk of triggering negative reinforcement~\cite{Zhai2024overreliance, passi2022overreliance, Ma2023llmwellbeing, chiu2024assessment}.

Research suggests that when individuals are given the freedom to make choices and exercise control over their reflective processes, they tend to experience deeper engagement driven by intrinsic motivation~\cite{patall2008effects}. This autonomy fosters a stronger sense of ownership over their personal growth, increasing their responsibility for and investment in the outcomes of their reflection~\cite{ryan2000self, deci2008self, patall2008effects, meng2015live, evans2015optimizing}. Recently in HCI, self-guided tools have been studied in the context of therapeutic intervention (\eg{}, \cite{sharma2023reframing, sharma2024facilitating}), allowing individuals to engage with LLM-driven systems to reframe negative patterns of thinking. These systems aim to improve the accessibility and effectiveness of interventions by guiding users through well-established tasks, such as identifying cognitive distortions and reframing, with the LLM providing adaptive support throughout the cognitively and emotionally demanding process. However, while such predefined therapeutic modules provide focused support, in this case on correcting cognitive distortions, these systems may still keep the users from exploring the freely flowing thoughts and feelings of users and coming to a resolution or new self-insight on their own. In this work, we are interested in designing an LLM-driven writing for well-being activity that can preserve user agency throughout the reflection process while still providing adaptive support. Striking this balance between guidance and autonomy is crucial, and we turn to formative studies where we explore ways to integrate adaptive LLM support within the context of users' reflective journeys.
\section{Formative Interviews}
To inform the design of \sysname{}, we conducted semi-structured interviews with mental health professionals, whose main role is to guide individuals in exploring and addressing their personal challenges. By understanding their underlying strategies and expertise, we aimed to gain a nuanced understanding of the challenges of providing effective guidance, which in turn informs the key considerations for designing an interface that incorporates therapeutic principles for its guidance, as well as to obtain insights into the potential and proper role of AI. 

\subsection{Procedure and Analysis}
\begin{table}[t]
\sffamily
\small
	\def\arraystretch{1.3}\setlength{\tabcolsep}{0.22em}
		    \centering
\caption[The table presents the demographic information of experts who participated in the formative study. It is structured into four columns: `Alias,' `Gender', `Job title,' `Years of experience,' and `Country of experience.' The table lists nine experts (E1 to E9) with their respective job titles, which include licensed counselors, clinical psychologists, and a psychiatrist. The gender distribution consists of eight females and one male. The years of experience for each expert range from 5 to 26 years. The experts have experience working in various countries, including South Korea, the United States, and the United Kingdom. The table provides a clear summary of the professional backgrounds of the experts involved in the study.]{Demographic information of the experts who participated in our formative study.}~\label{tab:experts-demographics}

\begin{tabular}{|c!{\color{gray}\vrule}l!{\color{tablegrayline}\vrule}m{0.31\columnwidth}!{\color{gray}\vrule}m{0.17\columnwidth}!{\color{tablegrayline}\vrule}m{0.23\columnwidth}|}
\hline
\rowcolor{tableheader}   \textbf{Alias} & \revised{\textbf{Gender}} & \textbf{Job title} & \textbf{Years of\newline{}experience} &\textbf{Country of\newline{}experience}\\ 
\hline
	\textbf{E1}  & \revised{Female} & Licensed counselor  & 16 years& South Korea             \\ \arrayrulecolor{tablegrayline}\hline
\textbf{E2}  & \revised{Female}  & Clinical psychologist  & 26 years & South Korea                     \\\hline
\textbf{E3}  & \revised{Female}  & Licensed counselor  & 15 years & South Korea                     \\\hline
\textbf{E4}  & \revised{Female}  & Licensed counselor  & 17 years & South Korea                     \\\hline
\textbf{E5}  & \revised{Female}  & Psychiatrist  & 18 years  & South Korea                     \\\hline
\textbf{E6}  & \revised{Female}  & Clinical psychologist  & 20 years  & United 
States,\newline{}United Kingdom                     \\\hline
\textbf{E7} & \revised{Female}   & Clinical psychologist  & 11 years  & United States                     \\\hline
\textbf{E8} & \revised{Male}  & Clinical psychologist  & 5 years & United States                     \\\hline
\textbf{E9} & \revised{Female}  & Clinical psychologist  & 8 years & United States                     \\
\arrayrulecolor{black}\hline
	
\end{tabular}
\end{table}

We recruited nine experts (E1--9\revised{, eight females and one male;} see \autoref{tab:experts-demographics}) through snowball sampling and internal network, who have extensive experience in clinical psychology and professional counseling. \revised{We initially planned to recruit experts in South Korea as our system was intended to be evaluated with the Korean population. However, the prolonged medical crisis in South Korea in 2024 made it difficult to recruit clinical psychologists locally. To supplement expert feedback from clinical psychologists, we recruited four clinical psychologists from outside South Korea through our internal network. The final nine participants} comprised five clinical psychologists, three licensed therapists, and one psychiatrist. Five were based in South Korea, three in the United States, and one had experience working across the United States and the United Kingdom. Participants had an average of 15.11 years (ranged 5 to 26) of experience in mental health care. 
Each interview lasted about one hour and was conducted in person or remotely, depending on the participants' availability. We compensated participants with 100,000 KRW or 100 USD, depending on their residence.

The interviews covered participants' strategies for helping clients navigate personal challenges, typical difficulties clients face, and potential roles they envision for LLM-driven systems in the therapeutic process. \revised{Since the experts have cared for various client populations in different settings, we asked them to share insights applicable to the general public.} To elicit their feedback, we \revised{also} presented the experts with an example personal challenge narrative \revised{(see Appendix \ref{app:formative-example})} and asked them to describe how they would approach such a case, including how they might guide the client through the process. \revised{Three researchers devised the example narrative during a brainstorming session, focusing on creating culturally neutral challenges.}
Additionally, we presented a paper prototype of an LLM-driven guided writing where the system suggests related topics based on the \revised{same example personal challenge narrative.} We asked for participants' feedback on its potential utility and areas for improvement.

All interviews were audio-recorded and transcribed for analysis. Employing thematic analysis~\cite{Braun2006ThematicAnalysis}, one researcher coded transcripts and grouped them into broader themes. The research team iterated several rounds of discussion to refine themes. In the following, we cover the findings from the interviews.

\subsection{Findings}
\ipstart{Importance of Guided Exploration in Addressing Individual Needs}
Participants highlighted the essential role of therapists in tailoring guidance to explore the unique needs of clients. E8 noted, ``\textit{When clients first enter therapy, they typically have a vague desire to `feel better' but struggle to articulate specific goals.}'' To address such ambiguity, therapists often start conversations to help clients clarify their objectives and understand the underlying issues they need to address. This process not only sets effective goals for the therapy but also empowers clients to gain a clearer understanding of their own difficulties. 
To guide their clients, all participants commonly used a strategy of asking a series of focused, open-ended questions to encourage self-reflection.
E2 shared that she often encourages clients to write about their issues between sessions, but many times, they struggle to decide what topic to write about. E2 remarked, ``\textit{I might suggest they write about certain things that arise from our collaborative exploration and give them specific assignments.}'' 
%

\ipstart{Challenges in Effective Questioning during Conversations}
All participants highlighted the challenges of asking adequate questions that the client would be ready to accept and explore. E6 noted, ``\textit{If I ask certain questions too early, the client might push back, thinking I'm making assumptions},'' emphasizing that clients should feel understood and supported, rather than overwhelmed or led in a direction they are not prepared to explore. Similarly, E4 mentioned, \textit{``My biggest challenge is with clients who don’t have much to say—they may shut down or resist answering. In these cases, I often have to find alternative questions or pathways to re-engage them.''}
Participants pointed out that there is no holy grail when it comes to the sequence of questions, and therefore, the effectiveness of any therapeutic approach depends largely on the therapist's ability to decide when and what the client would find it acceptable. 

\ipstart{Guidance to Enhance Awareness of the Process}
Participants emphasized the importance of keeping clients informed about the the status of the current therapeutic conversation, in part to stay focused and to prevent them from feeling overwhelmed or lost in their thoughts. 
They also noted that understanding the flow of the session and the broader discussions context helps clients construct coherent narratives and stay engaged in their own problems. For those struggling with ruminative thinking, where thoughts can circle back repeatedly without recognizing the broader context, these reminders become particularly important: E1 noted, \textit{``They often don't realize they're going in circles until I reflect back on what they have said, which helps them recognize the pattern,''}. E4 mentioned that she uses visual aids during sessions by writing down key themes or keywords as the conversation progresses.

\section{\sysname{}}
In this section, we first cover the design rationales we derived from prior literature and the formative study. We then describe the system design and user interfaces of \sysname{} and implementation details, including the generative pipelines.

\subsection{Design Rationales}

\ipstart{DR1. Provide diverse pathways to foster user agency and deep reflection} 
Recognizing that individuals may explore their thoughts through various cognitive pathways, it is crucial to foster flexible thinking and support a sense of autonomy in the reflective process~\cite{kashdan2010psychological, deci1987support, smallwood2013distinguishing}. In our formative study, professionals emphasized the importance of allowing clients to control their own reflective journeys. Therapists adjust their questions based on the client’s responses and collaboratively identify topics to prioritize for deeper exploration.
Therefore, we designed \sysname{} to provide a rich resource that enables users to take the lead in shaping and guiding their reflective pathways.
First, the \sysname{} generates ``Themes''---potential areas of reflection---based on the initial narrative provided by the user. Second, once the user selects a theme, the system suggests multiple ``Socratic questions''~\cite{clark2015socratic}, an evidence-based approach that encourages critical thinking and self-reflection. These questions prompt users to explore underlying emotions and perspectives that may not have surfaced in their initial narrative. 

\ipstart{DR2. Provide guidance to support expression without directing content}
In our formative study, professionals highlighted the usefulness of providing subtle guidance to help individuals articulate their thoughts and emotions. They emphasized that people often struggle with open-ended or thought-provoking questions, which can be cognitively overwhelming. 
In the context of writing for well-being, the benefit lies in the individual's process of translating their experiences into language~\cite{pennebaker1997writing}. To preserve this, our system avoids directly shaping the content of users' expressions (\eg{} generating direct sentences for users to adopt~\cite{kim2024diarymate}), instead offering scaffolding that gently supports the reflective process. To achieve this balance, we introduced features like ``Keywords'' and ``Comments.'' The keyword list offers users a subtle prompt by suggesting relevant concepts or terms that might help them think through a particular Socratic question, while still leaving room for their own interpretation. Additionally, the system generates comments that provide hints or feedback without making direct assertions about the user's narrative.

\ipstart{DR3. Provide a tangible summary of the progress}
In our formative study, professionals emphasized the importance of restating or summarizing client’s thoughts back to them, as this helps foster reflection by offering an objective perspective on the client’s experiences. They noted that while the summary might be subjective from the therapist’s viewpoint, it allows individuals to compare their internal thoughts with an external interpretation, serving as a valuable tool for stepping back and critically evaluating their journey. 
In line with this practice, we introduced the ``AI Summary'' feature, designed to offer users an objective summary of their reflections\revised{, distilled from} the user’s themes, questions, and answers, reflecting the key points of their exploration. 

\begin{figure}[t]
    \centering
    \includegraphics[width=\columnwidth]{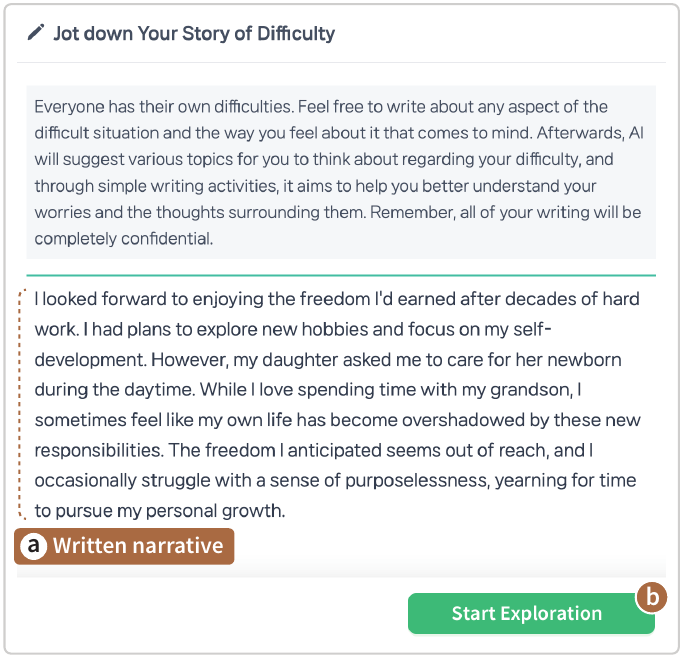}
    \caption[The figure shows the Initial Narrative page. Users are prompted to write their personal narrative (labeled `a'), which serves as a starting point for the system to generate adaptive guidance. A ``Start Exploration'' button (b) is present below the narrative input area, leading users to the Exploration page.]{\revised{The Initial Narrative page of \sysname{}. Writing the initial narrative~\circledigit{a} is a starting point for providing the basis clues to the system. Clicking the `Start Exploration' button~\circledigit{b} leads to the Exploration page (\autoref{fig:interface:components}-\blackrectsmall{A}).}}
    \label{fig:system:narrative}
\end{figure}

\begin{figure*}[t]
    \centering
    \includegraphics[width=\textwidth]{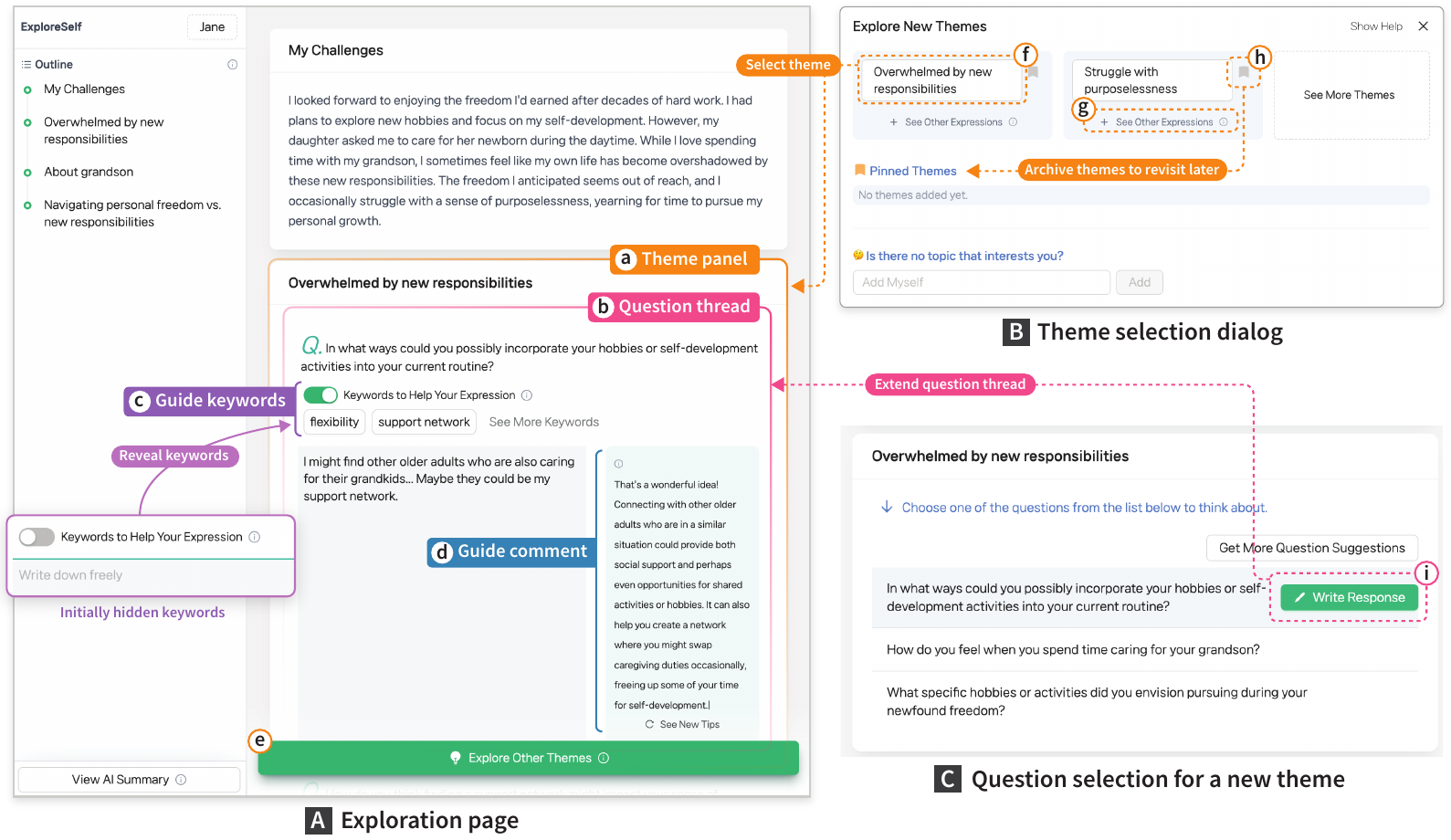}
    \caption[
    The image displays the user interface of \sysname{}, a system designed to support user-driven exploration and reflection on personal challenges. The interface consists of multiple interactive components: 
    1) On the left, the system provides an outline of themes such as ``My Challenges,'' ``Overwhelmed by new responsibilities,‘’ ``About grandson,‘’ and ``Navigating personal freedom vs. new responsibilities.‘’ 
    2) In the middle of the screen, the selected theme ``Overwhelmed by new responsibilities’' is open for exploration. This section shows a question thread where the user can reflect by responding to Socratic questions like ``In what ways could you possibly incorporate your hobbies or self-development activities into your current routine?‘’ 
    3) Users are also provided with AI-generated keywords like ``flexibility’' and ``support network,‘’ which can be toggled on or off (labeled as guide keywords). 
    4) Additionally, a guide comment is displayed, offering further reflection prompts, such as providing ideas on creating support networks. 
    5) On the right, the theme selection dialog shows the option to select or archive themes for later exploration. Multiple themes are suggested, including ``Overwhelmed by new responsibilities’' and ``Struggle with purposelessness.‘’
    6) Users have the option to extend the question thread or select new questions and explore new themes.
    The image visually highlights the core interactive elements such as themes, question threads, guide keywords, and comments, which are designed to facilitate an adaptive, user-driven reflective process.
    ]{\revised{Detailed interactions of the Exploration page~\blackrectsmall{A} of \sysname{}. The main scroll panel vertically appends a theme panel \circledigit{a} when the user creates a new theme ~\circledigit{f} on the Theme selection dialogue~\blackrectsmall{B}. The theme panel initially displays only the list of AI-suggested questions regarding the theme~\blackrectsmall{C}, and the user can populate the question thread~\circledigit{b} by selecting the next question they want to answer~\circledigit{i}. While answering a question, users can reveal keywords and request more~\circledigit{c}. The system also provides a guide comment~\circledigit{d}, which can be regenerated upon request.}}
    \label{fig:interface:components}
\end{figure*}

\begin{figure*}[t]
    \centering
    \includegraphics[width=\textwidth]{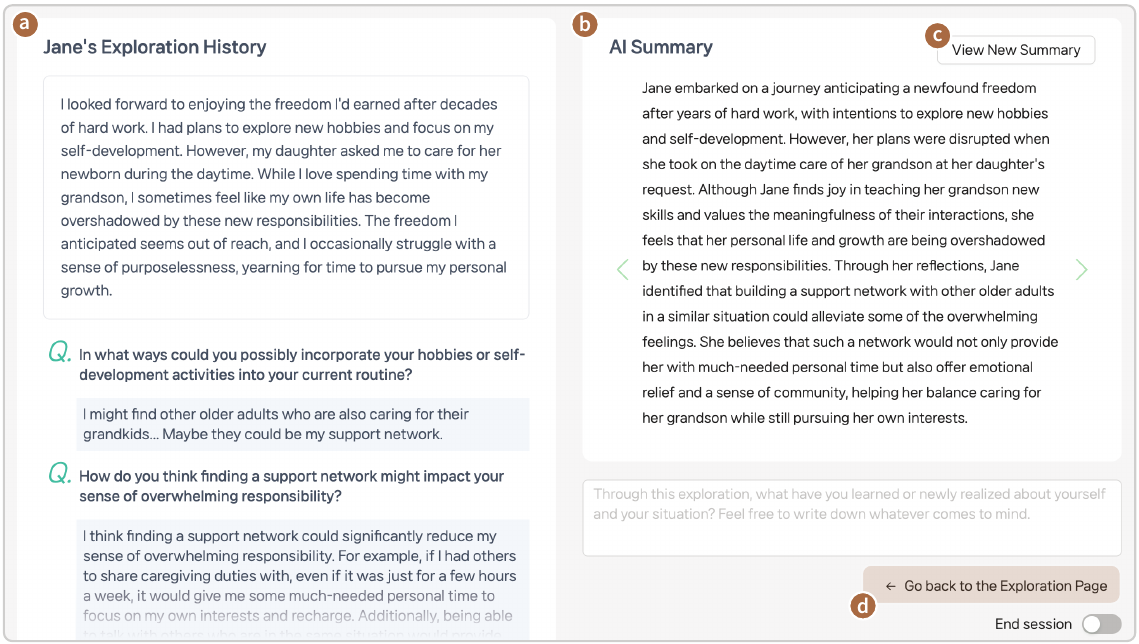}
    \caption[The figure shows the AI Summary page. Users can view their exploration history (a), including both their initial narrative and responses to questions. The AI-generated summary (b) provides a summary of their reflections, and users can generate a new summary by clicking the ``View New Summary'' button (c). There's also an option to navigate back to the Exploration page.]{\revised{The AI Summary page of \sysname{} where the users can overview their exploration history~\circledigit{a} and the AI-generated summarization~\circledigit{b}. The `View New Summary' button~\circledigit{c} generates a new summary. Users can go back to the Exploration page~(\autoref{fig:interface:components}-\blackrectsmall{A}) using the `Go back' button~\circledigit{d}.}}
    \label{fig:system:summary}
\end{figure*}

\subsection{System Design and User Interface}
\sysname{} is designed to assist users in navigating, exploring, and making sense of personal challenges through writing with structured and adaptive guidance by an LLM. The application consists of three primary phases of user engagement: (1) \textit{Writing the initial narrative} \revised{(\autoref{fig:system:narrative})}, (2) \textit{Exploring the narrative} \revised{(\autoref{fig:interface:components})}, and (3) \textit{Wrapping up and summarization} \revised{(\autoref{fig:system:summary})}. Each phase was carefully designed to encourage different aspects of self-reflection on and understanding of the personal challenges, helping users process their thoughts and emotions effectively. In the following, we describe each phase and interactions through a usage scenario: \textit{\exampleusername{} is a retiree who looked forward to enjoying the freedom she earned after decades of hard work. She had plans to explore new hobbies and focus on self-development. However, her daughter asked \exampleusername{} to care for her newborn during the daytime. Although \exampleusername{} loves spending time with her grandson, she sometimes feels that her own life has become overshadowed by her new responsibilities. The freedom she anticipated seems out of reach, and she occasionally struggles with a sense of purposelessness, yearning for time to pursue her personal growth. Now that she turns to \sysname{} to process this challenge.}

\subsubsection{Writing the Initial Narrative}
\exampleusername{} opens \sysname{} and the web page shows the \textit{Initial Narrative} page (\autoref{fig:system:narrative}), which prompts her to articulate her challenges. The interface presents a borderless text area encouraging \exampleusername{} to write freely about what weighs on her mind and her thoughts and feelings associated with it (\circledigit{a} in \autoref{fig:system:narrative}). 
\exampleusername{} jots down her story and clicks the `Start Exploration' button (\circledigit{b} in \autoref{fig:system:narrative}), which sends her to the Exploration phase, where the system facilitates a more structured process of introspection and analysis.

\subsubsection{Exploring the Narrative}
The \textbf{Exploration} page (\autoref{fig:interface:components}-\blackrectsmall{A}) allows \exampleusername{} to explore and reflect on her challenges by navigating both the breadth and depth of her narrative.

\ipstart{Themes: Expanding breadth} 
\exampleusername{} clicks the `Explore Themes' button (\circledigit{e} in \autoref{fig:interface:components}-\blackrectsmall{A}), and it opens the \textbf{Theme Selection} modal dialog (\autoref{fig:interface:components}-\blackrectsmall{B}). The dialog lists two AI-generated themes related to her initial narrative: `\texttt{Overwhelmed by new responsibilities}' and `\texttt{Struggle with purposelessness}.' She would like to explore both themes but saves the second one for later by clicking the bookmark icon (\circledigit{h} in \autoref{fig:interface:components}-\blackrectsmall{B}), which stores the theme in \textbf{Pinned Themes}. She then clicks the `\texttt{Overwhelmed by new responsibilities}' theme (\circledigit{f} in \autoref{fig:interface:components}-\blackrectsmall{B}). And a new \textbf{Theme panel} (\autoref{fig:interface:components}-\blackrectsmall{C}) appears on the main page.


\ipstart{Question Threads: Expanding depth} 
The theme panel lists three AI-generated Socratic questions that may encourage deeper reflection on the current theme. Of these questions \exampleusername{} selects the first one: ``\texttt{In what ways could you possibly incorporate your hobbies or self-development activities into your current routine?}'' (\circledigit{i} in \autoref{fig:interface:components}-\blackrectsmall{C}) creating a new \textbf{Question panel} (\circledigit{b} in \autoref{fig:interface:components}-\blackrectsmall{A}) on the theme panel. The question panel contains a text field to write down answers. Next to it, an \textbf{AI comment} (\circledigit{d} in \autoref{fig:interface:components}-\blackrectsmall{A}) is generated, encouraging \exampleusername{} to think about small pockets of time. To get more guidance for writing answers, \exampleusername{} turns on the switch for the \textbf{Keyword list} (\circledigit{c} in \autoref{fig:interface:components}-\blackrectsmall{A}) above the text field, which would reveal two relevant keywords, `\texttt{flexibility}' and `\texttt{support network}.' Inspired by the support network keyword, \exampleusername{} jots down the idea of finding peer older adults who also care for grandkids to form a support network. She clicks the `See New Comment' button (\circledigit{d} in \autoref{fig:interface:components}-\blackrectsmall{A}) to refresh the comment, and the new comment acknowledges her answer. \exampleusername{} goes on this thread of questions; she clicks the `Get More Question Suggestions' button (\autoref{fig:interface:components}-\blackrectsmall{C}) on the current question panel, and the Theme panel generates three new questions based on her question and answer. She selects `\texttt{How do you think finding a support network might impact your sense of overwhelming responsibility?}' that directly follows up her previous thread. She continues exploring the theme by creating and answering AI-driven questions.

\npstart{} Satisfied with the exploration of the ideas of the support network, \exampleusername{} turns to a new theme. She again clicks the `Explore Other Themes' button, and the Theme selection dialog suggests new themes generated considering her activities so far. However, no suggested themes inspire her. So \exampleusername{} \textbf{creates her own theme}, ``\texttt{About grandson},'' by typing it on a text field on the Theme selector dialog, to reflect on her grandson. The new Theme panel suggests questions related to her custom theme, and \exampleusername{} selects the question, ``\texttt{What are the aspects of spending time with your grandson that bring you joy and fulfillment?}'' Upon this question, her perspectives on caring for her grandson transitioned from being overshadowed to being fulfilled. She continues exploring through her challenging story by adding themes and questions. She also goes back to see past themes and her writing by clicking themes on the \textbf{Overview} list (\autoref{fig:interface:components}-\blackrectsmall{A}, top-left). 
After a while of exploration, \exampleusername{} feels that her exploration is saturated. She clicks the `View AI Summary' button (\autoref{fig:interface:components}-\blackrectsmall{A}, bottom-left) to wrap up the exploration.



\subsubsection{Wrapping Up and Summarization}
Jane enters the \textbf{AI Summary} screen (\autoref{fig:system:summary}). On the left column, she can review a comprehensive list of themes, questions, and answers (\circledigit{a} in \autoref{fig:system:summary}). On the right column, an AI-generated summary is displayed (\circledigit{b} in \autoref{fig:system:summary}). The summary \revised{distills} the key themes and insights from the user’s \revised{text} into a concise essay. Using the back button (\circledigit{d} in \autoref{fig:system:summary}), she can also go back to the Exploration screen and return with updated exploration paths. \exampleusername{} reads through the AI summary and ends the exploration.

\subsection{Generative Pipelines}
\begin{figure*}
    \centering
    \includegraphics[width=\textwidth]{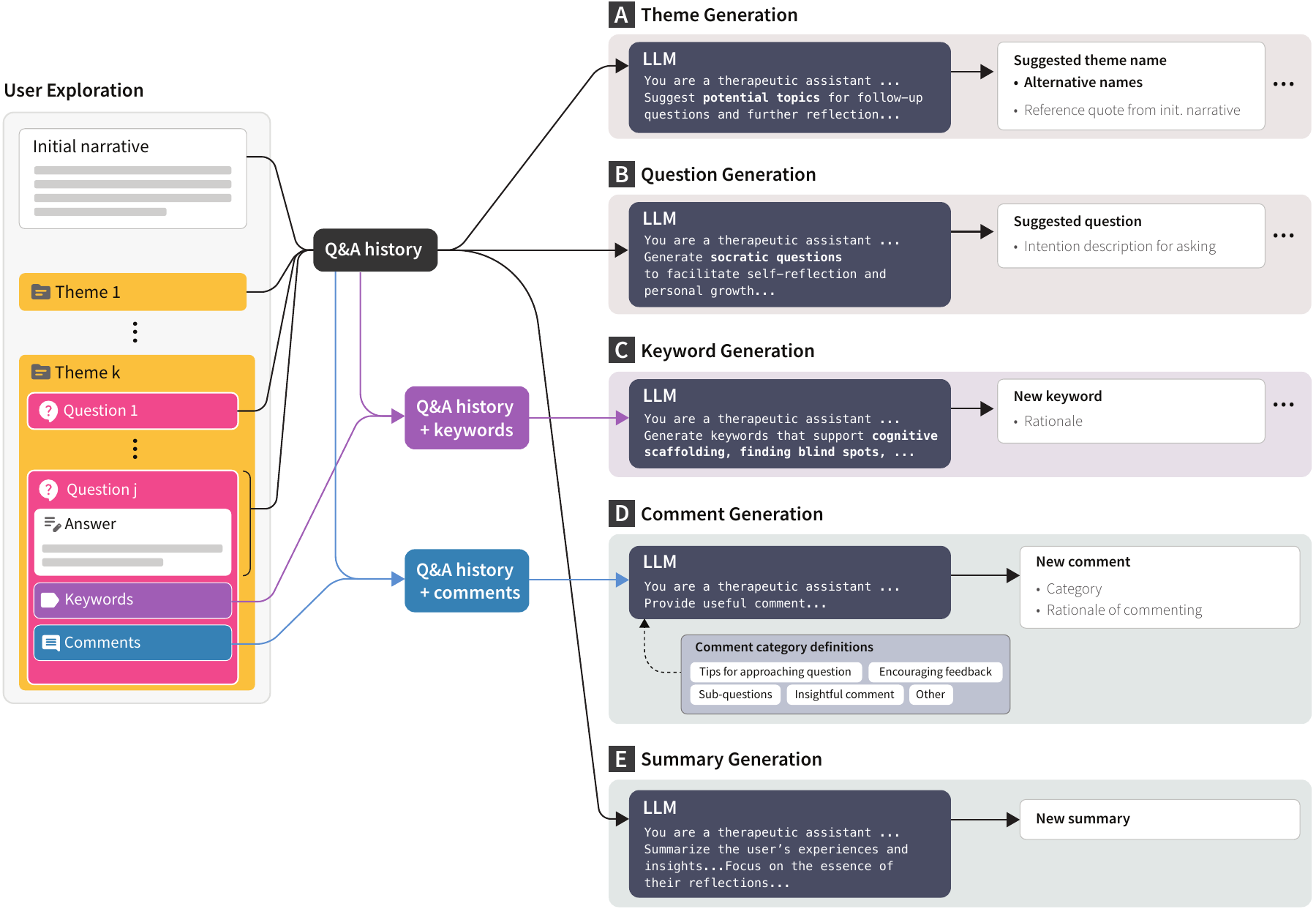}
    \caption[The figure illustrates the generative pipelines in \sysname{} that create themes, questions, keywords, comments, and summaries. On the left side, a user’s exploration, starting from the initial narrative, flows into several elements like themes, questions, and answers with keywords and comments. The Q\&A history feeds into different generative pipelines, each driven by an LLM. In pipeline A, the LLM generates suggested themes and alternative names based on the user’s narrative. In pipeline B, the LLM generates Socratic questions intended to encourage self-reflection. Pipeline C focuses on generating keywords that offer cognitive scaffolding, while pipeline D generates comments categorized as tips, feedback, or insights, with rationales for each comment. Pipeline E summarizes the user’s exploration, consolidating their reflections into a new summary. Each pipeline is guided by chain-of-thought style instructions to enhance the reliability of the generated output.]{The overview of pipelines to generate themes~\blackrectsmall{A}, questions~\blackrectsmall{B}, keywords~\blackrectsmall{C}, comments~\blackrectsmall{D}, and the summary \revised{of} the exploration history~\blackrectsmall{E} in the middle of exploration. Each generative pipeline incorporates an LLM inference that receives the current exploration history starting from the initial narrative as an input. The LLM inference is driven by chain-of-thought style instructions, providing descriptions and rationales for each output to enhance reliability.}
    \label{fig:system:pipelines}
\end{figure*}

In this section, we describe the generative pipelines incorporated in \sysname{} to support the generation of themes, questions, keywords, comments, and summaries. \autoref{fig:system:pipelines} illustrates the data flow and a gist of LLM instructions for each pipeline. The pipelines refer to the current exploration status (see \autoref{fig:system:pipelines}, left), including the initial narrative, themes, questions and answers, and keywords and comments selectively depending on the type of pipeline. The exploration status is inserted to the structured input instruction formatted in XML. The underlying LLMs run on instructions following the chain-of-thought prompting approach~\cite{wei2022chain}, where the model is instructed to provide meta-output (see thin texts in \autoref{fig:system:pipelines}, right), such as rationales of the output, along with the requested output to enhance the reliability of generation. We consistently applied a `therapeutic assistant' persona to the instructions (see dark boxes in \autoref{fig:system:pipelines}, right) to better contextualize the generation~\cite{wei2023leveraging, ha2024clochat}.


The \textbf{Theme Generation} step (\autoref{fig:system:pipelines}-\blackrectsmall{A}) identifies themes that arise from their personal narratives and previous interactions with the system. The model is instructed to name themes that closely align with the user's language and expressions. In case the user does not grasp the meaning of the original theme name, the model also provides \textit{alternative expressions} to allow the user reveal them on the Theme selector dialog (see \circledigit{g} in \autoref{fig:interface:components}-\blackrectsmall{B}). To further guide this behavior, the model clarifies which part of the initial narrative each theme stems from (see the white box in \autoref{fig:system:pipelines}-\blackrectsmall{A}) (\cf{},~\autoref{app:theme-gen}). The \textbf{Question Generation} step (\autoref{fig:system:pipelines}-\blackrectsmall{B}) suggests probing questions for a given theme. The model is instructed to craft socratic questions~\cite{clark2015socratic} that encourage users to navigate deeper into their thoughts and feelings. Each question is designed to resonate with the user’s language and context (\cf{},~\autoref{app:question-gen}).

To further guide the user while reflecting on the questions, the system also provides \textbf{keywords} (\autoref{fig:system:pipelines}-\blackrectsmall{C}) and \textbf{comments} (\autoref{fig:system:pipelines}-\blackrectsmall{D}). 
When generating keywords, the model is instructed that they should (1) serve as cognitive scaffolding, (2) activate potential blind spots for the user, and (3) be relevant to the user's background and core values (\cf{},~\autoref{app:keyword-gen}).
As for the comments, the model is instructed to first decide the category of the comment from one of the predefined categories: (1) tips for approaching the question, (2) encouraging feedback, (3) sub-questions elaborated from the questions, (4) insightful comment, and (5) others (See~\autoref{app:comment-gen}).

When generating \textbf{AI summary} (\autoref{fig:system:pipelines}-\blackrectsmall{E}), the pipeline \revised{distills} the user's themes, questions, and answers into an essay in several paragraphs. The model is instructed to generate a concise and focused story, using the user’s own language and expressions where appropriate. The summary is intended to highlight key themes, emotions, and notable progress, offering a clear snapshot of the user’s pathway of exploration. Also, the model tries to make the summary commensurate with the volume of the content of the user's writing, not extrapolating details and staying grounded in the user’s actual input.  \revised{In particular, the model prompt (\cf{},~\autoref{app:summary-gen}) includes guidelines ``\texttt{Use the user’s own language and expressions where appropriate}'' and ``\texttt{Keep the summary realistic, proportional to the content of the user’s log, and grounded to user’s input}.''} 

\subsection{Implementation}
\sysname{} consists of (1) a backend server that manages user data and LLM pipelines and (2) a frontend web interface, both written in TypeScript~\cite{TypeScript}. The server runs on the Express.js~\cite{Express} framework, providing core functionalities via REST APIs and storing the user data in MongoDB~\cite{MongoDB}. We implemented the LLM pipelines using LangChain.js~\cite{LangChain} incorporating OpenAI~\cite{openai}'s GPT-4 ChatCompletion API. As LLM's capabilities of dealing with non-English texts are inferior to equivalent English texts due to the inefficient tokenization~\cite{petrov2023tokenizerUnfairness}, we chose \texttt{gpt-4o}\footnote{\url{https://openai.com/index/hello-gpt-4o/}} as an underlying model, which involved an improved Korean tokenizer with state-of-the-art performance. The web interface was implemented using React~\cite{React}.
\section{User Study}
We conducted an exploratory study to examine how individuals interact with \sysname{} and how it may foster exploration and reflection on personal challenges. \revised{As part of the study, participants explored their own challenges using \sysname{} on their computers.} The study was conducted online via Zoom \revised{to allow participants to engage from a comfortable and secure} environment of their choice\revised{. This online approach enabled participants} to interact with \sysname{} at their own pace and in their preferred settings, which \revised{has been shown to foster} more authentic and reflective responses~\cite{schmader2018state, hoogesteyn2020detainee}. 
\revised{We refined both the system and the study protocol through three pilot sessions with Korean adults. For example, we incorporated the Pinned Themes feature, enabling users to pin and revisit previously generated themes. We also revised the wording used in the interface and the tutorial procedure to clarify that the system is not intended to provide solutions or counseling but to support thematic exploration.}
The study protocol was approved by the public institutional review board of the Ministry of Health and Welfare of South Korea. 

\subsection{Participants}

\begin{table}[t]
\sffamily
\small
	\def\arraystretch{1.1}\setlength{\tabcolsep}{0.5em}
		    \centering
\caption[The table presents demographic information of the participants in the study and their self-reported frequency of using AI assistants like ChatGPT and Bard. It is structured into five columns: 'Alias', 'Age', 'Gender', 'Occupation', and 'Frequency of AI assistant use'. The table lists 19 participants (P1 to P19), with ages ranging from 22 to 65 years, representing various occupations such as full-time professionals, graduate students, self-employed individuals, and homemakers. The reported frequency of AI assistant use varies across participants, with some using them daily (e.g., P1, P3, P5) while others report never having used them (e.g., P7, P10, P19). The table provides a detailed overview of the participant demographics and their interaction with AI technologies.]{Demographic information of the participants and their self-report frequency of using AI assistants like ChatGPT~\cite{chatgpt} and Bard~\cite{Bard}.}~\label{tab:demographic}

\begin{tabular}{|c!{\color{gray}\vrule}c!{\color{tablegrayline}\vrule}l!{\color{tablegrayline}\vrule}m{0.25\columnwidth}!{\color{tablegrayline}\vrule}m{0.31\columnwidth}|}
\hline
\rowcolor{tableheader}   \textbf{Alias} & \textbf{Age} & \textbf{Gender} & \textbf{Occupation} & \textbf{Frequency of\newline{}AI assistant use}\\ 
\hline
	\textbf{P1}    & 39  & Female  & Full-time (IT)    & At least once a day             \\ \arrayrulecolor{tablegrayline}\hline
\textbf{P2}    & 28  & Female & Graduate student  & Within 3 times a week                      \\\hline
\textbf{P3}    & 28  & Male & Unemployed  & At least once a day                   \\\hline
\textbf{P4}    & 44  & Male  & Self-employed & Within 3 times a week                          \\\hline
\textbf{P5}    & 26  & Male & Graduate student   & At least once a day            \\\hline
\textbf{P6}    & 44  & Female & Self-employed & At least once a day                \\\hline
\textbf{P7}    & 25  & Male   & Homemaker & Never                                  \\\hline
\textbf{P8}    & 53  & Female & Part-time (Sales)  & Tried once or twice                 \\\hline
\textbf{P9}    & 27  & Male     & Self-employed    & 4-6 times a week\\\hline
\textbf{P10}   & 41  & Female  & Full-time (Energy)   & Never                     \\\hline
\textbf{P11}   & 36  & Male  & Full-time\newline{}(Finance)   & Tried once or twice              \\\hline
\textbf{P12}   & 22  & Male   & Undergraduate student     & At least once a day     \\\hline
\textbf{P13}   & 37  & Female    & Preschool teacher   & Tried once or twice    \\\hline
\textbf{P14}   & 65  & Female  & Part-time\newline{}(Elderly welfare)  & At least once a day  \\\hline
\textbf{P15}   & 40  & Female   & Adjunct instructor     & 4-6 times a week         \\\hline
\textbf{P16}   & 32  & Female  & Full-time\newline{}(Healthcare)   & Tried once or twice             \\\hline
\textbf{P17}   & 39  & Male   & Private academy\newline{}instructor    & 4-6 times a week     \\\hline
\textbf{P18}   & 38  & Female    & University lecturer    & Within 3 times a week             \\\hline
\textbf{P19}   & 49  & Male     & Full-time\newline{}(Distribution)    & Never               \\ \arrayrulecolor{black}\hline

\end{tabular}
\end{table}

We recruited 19 participants (P1--19; 10 female) by advertising the study on a local social community platform and online university communities in South Korea.  
Our inclusion criteria were adults who (1) are 19 years old or older; (2) are native Korean speakers; (3) are currently staying in Korea; (4) can join an online study in a private and comfortable place; and (5) are motivated to better understand their personal challenges. In particular, we ensured participants met criteria (2) and (3), (2) for them to be able to express their personal challenges in the most comfortable language as \sysname{} was implemented in Korean. (3) was to make sure the participants were in an accessible proximity, allowing them to engage in the study from a private space of their choosing, where they could write without researchers physically present while still being accessible for the study. Please refer to \autoref{sec:study:safety} for more detail. The advertisement included a link to a screening survey that asked basic demographic information (\ie{}, age, gender, educational level, employment details, as well as background questions about prior experience with AI and personal endeavor to process personal challenges, if any.
\revised{We did not screen participants for the presence of mental illness, as \sysname{} was designed to support the general population's mental wellbeing rather than serve as a psychotherapeutic intervention.}

\autoref{tab:demographic} shows the demographic composition of our study participants. Participants were from 22 to 65 years old ($avg = 37.53$). The participants comprised 11 full-time or part-time workers in various sectors, including distribution and elderly welfare, three self-employed local business owners, and the rest consisting of two graduate students, an undergraduate student, an unemployed person, and a homemaker. Fifteen out of 19 participants had tried conversational AI assistants like OpenAI's ChatGPT and Google Bard (now Gemini) more than once in their daily lives, and six of them used AI assistants for general purposes at least once a day. Regarding the efforts to process personal challenges, a majority of participants (13 out of 19; 68\%) had explored books or materials like YouTube videos about mental health, seven participants had had professional counseling, and five participants had tried AI chatbots like ChatGPT and Bard to discuss their challenges. To compensate for their participation, we offered a 50,000 KRW (approx. 38 USD) online gift card.

\subsection{Safety Considerations}\label{sec:study:safety}
Although our target participant group did not specifically include individuals experiencing mental illness, we prepared safety protocols for the study based on the work of O’Leary et al.~\cite{o2018suddenly}, which involved peer support chat amongst individuals experiencing mental illness. This was a mindful procedure considering that \sysname{} encourages participants to disclose personal issues that might be sensitive and potentially traumatic, and participants could have mental health issues that were not disclosed to the experimenter.

 Prior to the study, participants were provided with a list of emergency contact numbers and municipal support resources to ensure they had access to immediate help if needed. During the study, participants were encouraged to take breaks at any time for any reason, upon notifying the researcher through a real-time messaging system if they required one. The first author carefully monitored participant interactions with \sysname{} to identify any potential risks, particularly focusing on signs such as mentions of self-harm or harm to others, as well as indications of increased distress, frustration, or exposure to triggering LLM-generated content. If any of these occurred, the first author would promptly engage with the participant to address the issue and take appropriate actions. These actions include immediately halting the session to secure participant safety; contacting municipal support resources on behalf of the participant in case of requiring help; and referring the participant to the healthcare center within the affiliated institution that included a consulting psychiatrist, should the participant needed to consult a clinician at once. We informed the participants of our partial compensation policy in case of any emergencies. In the course of the study progression with all 19 participants, no such event occurred.

\subsection{Study Setup and Procedure}
Each participant was invited to a 90-minute study session remotely via Zoom video call on their computer. Instructions for installing and using the remote sharing software were prepared in case of need. One researcher administered the sessions as an experimenter. The study session consisted of four parts: (1) briefing, (2) tutorial, (3) exploration with \sysname{}, and (4) debriefing.

\ipstart{Briefing} The experimenter shared her screen with presentation slides (refer to our supplementary material) and described the motivation and goal of our study. \revised{To encourage participants to authentically engage with the system, we informed them that the goal of this study is not to evaluate or address their specific personal challenges but to understand how our tool helps their exploration and reflection on those challenges.}

\ipstart{Tutorial} After explaining the goal of the study, the experimenter gave a 10-minute tutorial covering the main features of \sysname{} using a slide presentation. We masked out any dynamically generated or user-provided texts from the screenshots to minimize the potential bias from example contents that might influence participants' preconceptions of personal challenges and writing.

\ipstart{Exploration} After the tutorial, the experimenter shared a link to access \sysname{} and invited the participant to a Zoom breakout room, a virtual space where the participant could stay alone and focus. The experimenter was able to monitor the real-time interaction logs on \sysname{} to observe the activity patterns. The experimenter and the participant were connected via an instant messenger to communicate about any issues the participant may have during the exploration including help requests. The participant then freely used \sysname{} to explore their challenge. There was no minimum time requirement, and participants could finish the session whenever they felt they had sufficiently used it, for up to 45 minutes. At the 45-minute mark, the researcher checked in with participants if they needed more time. If so, they were allowed up to an additional 15 minutes, with a maximum session time of 1 hour. Since participants did not have free practice in the tutorial phase, the system offered on-demand help popups on major interface components such as the Theme Selector dialog. To assess the impact of the exploration on participants' attitudes towards their challenges, \revised{we asked participants to rate their perceived agency by completing the validated Korean version~\cite{choi2008validation} of the Pathways Subscale of the State Hope Scale~\cite{snyder1996development}, before and after of this phase. This 4-item measure asks participants to rate statements such as ``\textit{There are lots of ways around any problem that I am facing now}'' on an 8-point scale, ranging from 1 (Definitely false) to 8 (Definitely true). The Pathways Subscale of the State Hope Scale (refer to \autoref{app:pathway-scale}), used in this study, captures individuals' perceptions of their ability to achieve their goals~\cite{snyder1996development, schleider2020acceptability, buchlmayer2024effects}. Scores range from 4 to 32, with higher scores reflecting a stronger belief in the availability of pathways to their goals.}

\ipstart{Debriefing} At the end of the session, we removed the breakout room and conducted a semi-structured interview for about 30 minutes. Before starting the interviews, we confirmed their willingness to proceed. We asked participants about their experiences with \sysname{}, their rationales for adding and moving between themes and questions, receptiveness to AI-driven guidance, and how the interface helped or hindered their exploration and reflection.

\begin{table*}[b]
\sffamily
\small
\def\arraystretch{1.2}\setlength{\tabcolsep}{0.25em}
\centering
\caption[The table categorizes participants' initial narratives into five major categories: `Relationships', `Self-development', `Self-identity \& character', `Stability in life', and `Health'. It includes three columns: `Categories', `Description of categories', and `Sub-categories'. The categories represent different themes that emerged from participants' narratives, with multiple sub-categories assigned to each category. For example, under `Relationships' (reported by 12 participants), sub-categories include family conflicts, parenting, and societal pressure. `Self-development' (9 participants) covers academic stress, job dissatisfaction, and career uncertainties, while `Self-identity \& character' (8 participants) involves issues like aging, self-esteem, and existential concerns. Other categories include `Stability in life' (6 participants), which includes financial crises and life transitions, and `Health' (5 participants), covering mental and physical well-being challenges. The table is multi-coded, meaning each participant could contribute to multiple categories.]{The categorization of participants' initial narratives. The sub-categories are multi-coded to the initial narratives and grouped into categories. So, the number of participants for each category exceeds the total number of participants.}~\label{tab:narrative-categories}

\begin{tabular}{|m{0.18\textwidth}!{\color{gray}\vrule}m{0.29\textwidth}!{\color{gray}\vrule}m{0.5\textwidth}|}
\hline
\rowcolor{tableheader}   \textbf{Categories} & \textbf{Description of categories} & \textbf{Sub-categories}\\ 
\hline
\textbf{Relationships}\newline{}(12 participants) & Personal interactions and societal influences\newline{}affecting connections with others. & Family conflicts[P1, P8, P14, P15], parenting[P1, P7], social relationships[P6, P8], romantic relationships[P3, P16], societal stigma[P4], \newline{}societal pressure[P1, P11, P13, P19]
            \\ \arrayrulecolor{tablegrayline}\hline

\textbf{Self-development}\newline{}(9 participants) & Challenges in personal growth, career,\newline{}and managing goals. & Academic stress[P5], job dissatisfaction[P10], time-management concern,\newline{}goal fulfillment, career uncertainties[P1, P5, P6, P9, P12, P14, P15, P17]                    \\\hline
\textbf{Self-identity \& character}\newline{}(8 participants) & Issues related to personal identity,\newline{}self-perception, and existential concerns. & Aging[P8, P14], sexual identity[P3], existential concern[P8, P16],\newline{}self-esteem issue[P2, P7, P8, P13, P19]
                     \\\hline

\textbf{Stability in life}\newline{}(6 participants) & Challenges affecting financial \newline{}and personal stability. & Financial crisis[P8, P9, P11], financial instability[P5, P9, P18],\newline{}life transitions[P16]
\\\hline
\textbf{Health}\newline{}(5 participants) & Mental, emotional, and physical\newline{}well-being challenges. & Managing anxiety[P16, P18], trauma[P4, P8], emotional struggles[P2],\newline{}physical health concerns[P8]
                    \\
\arrayrulecolor{black}\hline
	
\end{tabular}
\end{table*}

\subsection{Data Analysis}
We employed both quantitative and qualitative methods to analyze our data, in order to gain a comprehensive understanding of participant engagement and interaction with \sysname{}. First, to understand what types of personal challenges participants would explore on our system, participants' initial narratives were categorized through an iterative open-ended coding process conducted by two authors. \revised{The first author, who has two years of experience in HCI and mental health research and administered the user study, first open-coded the initial narratives. As the initial narratives usually described multiple aspects, we multi-coded the narratives. Another researcher with a PhD in computer science and 12 years of experience in HCI and health informatics research participated in multiple rounds of discussion to finalize a codebook for sub-category labels. The full research team then iterated multiple sessions of discussions to group the subcategories into five broader categories. Although the coding was primarily bottom-up and grounded from the data, some category labels were informed by Sharma~\etal{}~\cite{sharma2023reframing}'s definitions of personal issues (\eg{}, \textit{parenting}, \textit{trauma}) and the life values set used in Acceptance and Commitment Therapy~\cite{hayes2005act}~(\eg{}, \textit{social relationships}). The categorization is reported in \autoref{tab:narrative-categories}.}

To analyze their overall usage and interaction patterns, We collected interaction logs to track key behaviors, such as the number of themes and questions selected by participants, the number of interactions with keywords and comments, and phase-switching behaviors (\eg{} between exploration and overview phase). These logs were analyzed to compute descriptive statistics, including the minimum, maximum, mean, and standard deviation for user engagement across different system features.

\revised{The Zoom sessions for debriefing interviews were recorded and transcribed later.}
We \revised{qualitatively} analyzed the debriefing transcripts \revised{using} thematic analysis~\cite{Braun2006ThematicAnalysis} to identify key themes regarding participants' experiences with \sysname{}. \revised{The first author open-coded the transcripts, going through several rounds of iterations. The full research team then discussed and identified patterns and themes through multiple rounds of peer debriefing. From this coding, we surfaced the key themes around the participants' sense of agency while using \sysname{} and how the system features supported the exploration, reflection, and self-expression.} 
\deleted{By combining quantitative metrics from interaction logs and qualitative insights from debriefing interviews, we explored how participants utilized the system features and how the system supported their reflective practices.}
\section{Results}
In this section, overall system usage based on quantitative metrics from interaction logs and surveys. We then cover how the system influenced participants' sense of agency during the exploration. Drawing on the qualitative analysis of the debriefing, we describe how themes and questions supported the user-directed navigation of personal challenges, and how keywords and comments helped participants' self-expression. Lastly, we report participants' perceived utility of \sysname{}.

\subsection{Overall System Usage}

Participants wrote about a wide range of personal challenges in their initial narratives, from personal relationships to dilemmas in life decision-making. \autoref{tab:narrative-categories} summarizes the categories of initial narratives participants entered. Participants mostly engaged with the challenges regarding \textit{personal relationships} (12 participants; 63\%), followed by \textit{self-development} (9 participants; 47\%), \textit{self-identity \& personal character} (8 participants; 42\%), \textit{stability in life} (6 participants; 32\%), and \textit{health} (5 participants; 26\%). The initial narratives often described multiple interconnected topics.

\begin{table*}[t]
\sffamily
\small
    \def\arraystretch{1.2}\setlength{\tabcolsep}{0.28em}
    \centering
\caption[The table summarizes participants' inputs, exploration, and guide usage during the study. It is structured into six metrics: syllables of the initial narrative, total syllables of responses, number of themes, total number of questions, total number of revealed keywords, and total number of comment requests. Each metric provides the mean and standard deviation, followed by detailed participant data (P1 to P19). For example, P1's initial narrative contains 233 syllables, and P2's narrative contains 478 syllables. Other data include the total syllables of responses, number of themes explored, questions answered, keywords revealed, and comment requests made. Some participants show high engagement with many keywords and comments, while others have minimal interactions with these features.]{Summary of syllable counts of the participants' inputs and the number of exploration and guide entities by participant. Note that for comments we counted only the comments that were manually requested by participants.}~\label{tab:usage-data}

\begin{tabular}{|l!{\color{gray}\vrule}c!{\color{tablegrayline}\vrule}c!{\color{gray}\vrule}c!{\color{tablegrayline}\vrule}c!{\color{tablegrayline}\vrule}c!{\color{tablegrayline}\vrule}c!{\color{tablegrayline}\vrule}c!{\color{tablegrayline}\vrule}c!{\color{tablegrayline}\vrule}c!{\color{tablegrayline}\vrule}c!{\color{tablegrayline}\vrule}c!{\color{tablegrayline}\vrule}c!{\color{tablegrayline}\vrule}c!{\color{tablegrayline}\vrule}c!{\color{tablegrayline}\vrule}c!{\color{tablegrayline}\vrule}c!{\color{tablegrayline}\vrule}c!{\color{tablegrayline}\vrule}c!{\color{tablegrayline}\vrule}c!{\color{tablegrayline}\vrule}c!{\color{tablegrayline}\vrule}c|} 
\hline
\rowcolor{tableheader} \textbf{Metric} & \textbf{Mean} & \textbf{SD} & \textbf{P1} & \textbf{P2} & \textbf{P3} & \textbf{P4} & \textbf{P5} & \textbf{P6} & \textbf{P7} & \textbf{P8} & \textbf{P9} & \textbf{P10} & \textbf{P11} & \textbf{P12} & \textbf{P13} & \textbf{P14} & \textbf{P15} & \textbf{P16} & \textbf{P17} & \textbf{P18} & \textbf{P19} \\ 
\hline
Syllables of initial narrative & 233.63 & 299.98 & 233 & 478 & 39 & 215 & 313 & 445 & 42 & 1347 & 100 & 180 & 198 & 13 & 137 &1,534 & 25 & 34 & 111 & 239 & 127 \\ 
\arrayrulecolor{tablegrayline}\hline
Total syllables of responses & 745.47 & 349.65 & 1,261 & 568 & 699 & 388 & 1,273 & 1,019 & 541 & 1,278 & 571 & 325 & 381 & 496 & 1,381 & 417 & 824 & 508 & 864 & 465 & 905 \\ 
\hline
\# of themes & 4.89 & 2.26 & 7 & 3 & 2 & 2 & 9 & 5 & 5 & 5 & 6 & 4 & 3 & 4 & 5 & 6 & 4 & 11 & 3 & 4 & 5 \\ 
\hline
Total \# of questions & 11.47 & 7.28 & 17 & 7 & 18 & 3 & 11 & 4 & 15 & 6 & 5 & 4 & 8 & 28 & 12 & 9 & 10 & 12 & 22 & 4 & 23 \\ \hline
Total \# of revealed keywords & 11.79 & 8.69 & 0 & 9 & 0 & 27 & 6 & 11 & 29 & 18 & 8 & 9 & 24 & 15 & 4 & 18 & 20 & 4 & 6 & 9 & 7 \\\hline
Total \# of comment requests & 6.47 & 8.26 & 0 & 1 & 0 & 2 & 0 & 5 & 18 & 4 & 1 & 0 & 6 & 1 & 11 & 3 & 10 & 27 & 9 & 1 & 24 \\
\arrayrulecolor{black}\hline
\end{tabular}
\end{table*}

\autoref{tab:usage-data} summarizes the syllable counts of participants' written inputs (\ie{}, initial narratives, and responses to questions) and the number of themes, questions, keywords, and comments generated or requested by participants. The average syllable count of the initial narratives was 233.63 with high variance~(\reportstats{299.96}{13}{P12}{1347}{P8}). The total syllable count of all responses per participant was 745.47 on average~(\reportstats{349.65}{325}{P10}{1381}{P13}).
There was a moderate correlation between the syllable count of the initial narrative and the total syllable count of the responses, with a Pearson correlation coefficient of 0.49 (\pval{0.03}), indicating a statistically significant relationship.

\begin{figure}[b]
    \centering
    \includegraphics[width=\columnwidth]{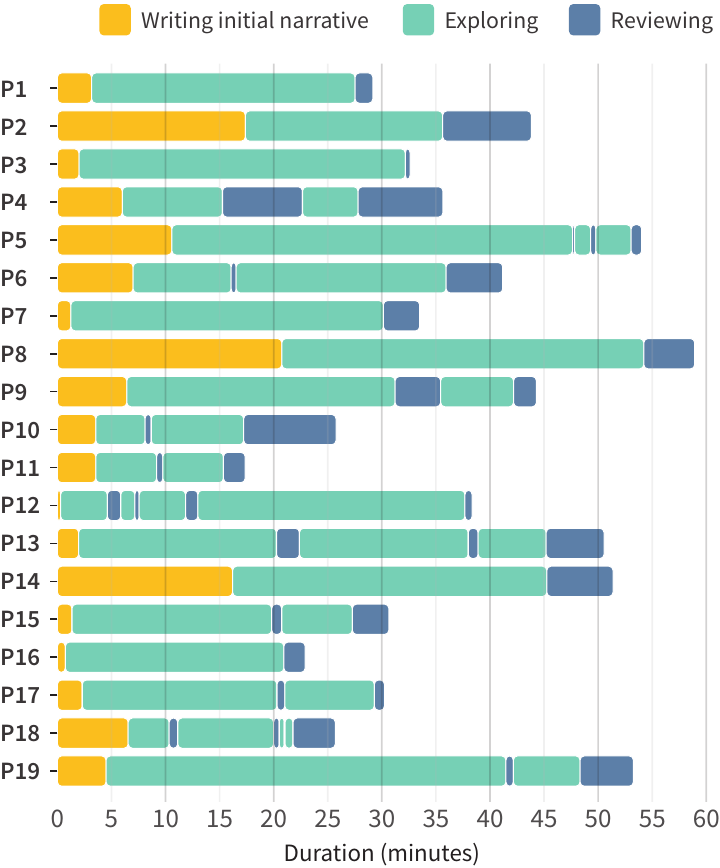}
    \caption[The chart shows the timelines of 19 participants (P1 to P19) during the exploration phase of the study, with time measured in minutes along the X-axis (ranging from 0 to 60 minutes). Each participant's timeline is divided into three color-coded segments: yellow bars indicate the time spent writing the initial narrative, cyan bars represent the time spent in the exploration phase, and blue bars denote the time spent reviewing the AI summary. The duration of engagement varies between participants, with some showing more exploration time, while others spent significant time reviewing the AI summary or writing the initial narrative.]{The timelines of participants during the exploration phase. The X-axis indicates elapsed minutes. The yellow bars denote the periods spent writing the initial narratives in the Initial Narrative page (\autoref{fig:system:narrative}), the cyan bars denote the periods staying in the Exploration page (\autoref{fig:interface:components}-\blackrectsmall{A}), and the blue bars denote the periods staying in the AI Summary page (\autoref{fig:system:summary}). Note that participants were able to move between the Exploration and the AI Summary pages.}
    \label{fig:results:durations}
\end{figure}

There was also a high individual variance in how participants spent time across screens. \autoref{fig:results:durations} illustrates the participants' timelines of the exploration phase, split by the type of screens they stayed on. None of the participants took breaks during their engagement, and they took an average of 37.86 minutes in the entire exploration phase~(\reportstats{11.58}{17.37}{P11}{58.93}{P8}). While writing the initial narrative, participants spent an average of 5.82 minutes~(\reportstats{5.40}{0.3}{P12}{17.38}{P2}; see yellow bars in \autoref{fig:results:durations}). Then they spent an average of 31.77 minutes~(\reportstats{9.42}{13.82}{P11}{48.8}{P13}) for both exploring and reviewing the AI summaries (see green and blue bars in \autoref{fig:results:durations}). 

\begin{figure}
    \centering
    \includegraphics[width=\columnwidth]{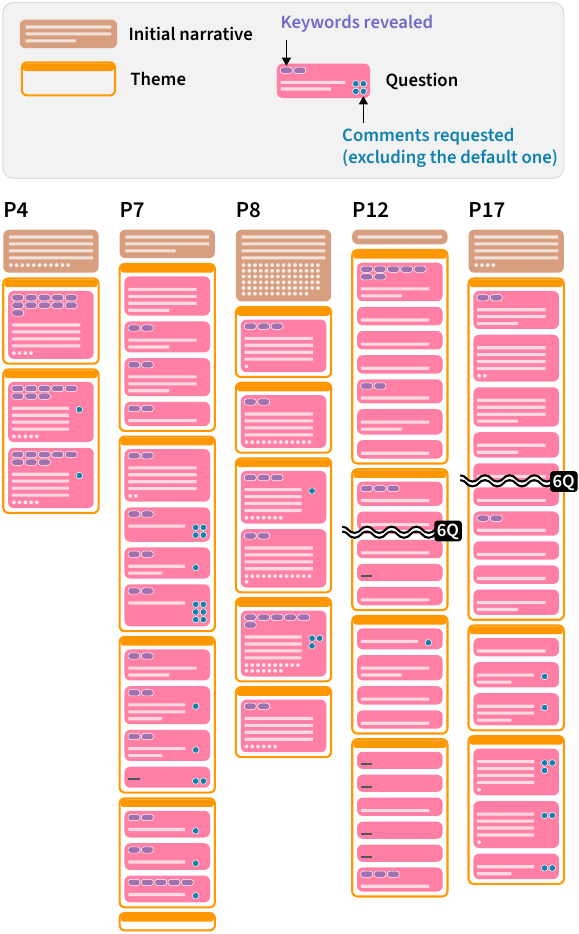}
    \caption[The figure displays a side-by-side comparison of the exploration patterns of five selected participants (P4, P7, P8, P12, and P17) in the study. Each participant’s activity is represented vertically. At the top of each participant’s sequence is the initial narrative, followed by a series of themes (yellow framed boxes), and within each theme, questions (pink boxes) are listed. Keywords revealed during the exploration are shown as purple capsules within the question boxes, and any additional comments requested are marked with blue dots. The size of the narrative and answers is visually represented by lines and dots, with each line and dot indicating 15 syllables. This structure illustrates how each participant engaged with the system's features, such as selecting themes, responding to questions, and requesting additional comments, offering insight into their reflective processes and patterns of engagement.]{The results of the exploration of selected participants with distinct patterns. The brown boxes denote the initial narratives, the yellow frames denote themes, and the pink boxes denote questions. Number of keywords that participants revealed is depicted as purple capsules. The blue dots denote the comments that participants additionally requested. The white lines and dots illustrate the amount of text for the initial narratives and answers (the lines and dots both denote 15 syllables.)}
    \label{fig:results:flow_participants}
\end{figure}

Participants showed diverse patterns in exploration, engaging with a varying number of themes and questions suggested by \sysname{}. \autoref{tab:usage-data} reports the descriptive statistics of participant engagement. On average, participants engaged with 4.89 themes~(\reportstats{2.26}{2}{P3, P4}{11}{P16}) and 11.47 questions~(\reportstats{7.28}{3}{P4}{28}{P12}). The number of questions per theme averaged 2.58~(\reportstats{2.61}{1}{P4, P5, P6, P8, P9, P10, P11, P13, P14, P16, P18}{16}{P17}). The total syllable count of responses written during the exploration phase averaged 796.95~(\reportstats{464.44}{269}{P12}{1674}{P13}). 
\autoref{fig:results:flow_participants} illustrates the results of exploration captured from selected participants, who showed distinct engagement patterns with different components. For example, P4 actively leveraged keywords (purple capsules in \autoref{fig:results:flow_participants}), and P7 generally used both the keywords and comments (blue dots in \autoref{fig:results:flow_participants}) throughout all questions. P8 wrote an extreme amount of text for the initial narratives and answers, and P12 showed a tendency to write a short narrative and answers. Lastly, P17 dove into the first theme with many questions.



\subsection{Sense of Agency over Exploration}

 Participants' perceived agency on their challenge increased after using \sysname{}: The pathway subscale score, which we collected before and after the system use, significantly increased from 22.32 ($SD=4.91$) to 24.95 ($SD=5.86$) by 2.63 on average ($t(18)=2.80$, \pval{0.012^{*}}). Cohen's $d$ of the two means was 0.66, indicating the effect size of the gaps falling between medium (0.5) and large (0.8). Specifically, the scores of 15 out of 19 (79\%) participants increased.

In debriefing, participants expressed varying perceptions of their agency while using \sysname{}, with several factors influencing their sense of control.  
Some participants felt that the \textbf{acknowledgment of their wordings in the AI-generated text} played a crucial role in shaping the experience. P11 observed that while the AI sometimes paraphrased his inputs into different expressions in its generation, this paraphrasing gave him the impression that the AI was taking an active role. However, he still felt a sense of agency because the AI-generated texts were grounded in his original wording, highlighting that his expressions ultimately drove the interaction. P8 echoed this sentiment, noting that the system’s ability to suggest themes based on her input made her feel that she was leading the process. 

Participants also mentioned that the \textbf{generation latency} of AI also influenced their sense of control. For instance, P14 highlighted that while she retained some control, \textit{``the AI’s faster processing of information made it feel more like the system was taking the lead at times, especially when [her own] thoughts formed more slowly.''}

The \textbf{maturity of reflections} further enhanced participants' perceived control over the process. P13 described how, in the early stages of each theme, she felt more guided by the AI, but as she began to clarify her thoughts, she began to feel more confident and in control in the theme thread, feeling that she was leading the exploration. This growing clarity allowed P13 to \textit{``confidently decide when to explore a theme more deeply and when to move on to a new one''} when she \textit{``felt [she] had fully explored the current one.''} This gradual shift from AI-driven to self-directed reflection was echoed by a few more participants who appreciated the system’s guidance in the initial phases but ultimately took ownership of the process as their reflections unfolded.

On the contrary, participants felt that their autonomy was compromised by the AI’s influence, especially when the AI’s suggestions of themes and questions derailed from their primary concerns and steered them toward unintended directions. P4 remarked, \textit{``The problem I originally had was related to my family, but the AI kept focusing on my band activities as if there were an issue there. Even though I knew that wasn't the case, I ended up revisiting the topic, which made the process take longer.''} 

In general, participants appreciated the way \sysname{} visualized and dynamically structured the breakdowns of their thoughts into a coherent landscape. Ten participants highlighted how this experience differed from their self-journaling practices. P18, for instance, shared that when reflecting on her own, her thoughts would often \textit{``spiral in a negative direction''} with no clear way to shift perspectives. In contrast, the system allowed her to \textit{``step back''} and revisit previous thoughts easily, helping her avoid getting stuck in a mental loop. 
Similarly, P7 remarked that if she had tried to write about her stress and various life challenges on her own, her thoughts would have \textit{``jumped from one issue to another,''} leading to a scattered reflection. However, with the AI’s structure, she could \textit{``stick to one theme''} at a time, making the process more focused and less overwhelming. 

\subsection{User-directed Navigation}

Participants navigated through themes and questions by making deliberate choices about whether to explore deeper into a particular theme (\ie{}, add a new question to the thread) or move on to a new theme. \deleted{Some participants managed their emotional engagement with a theme by deciding when to stop drilling into certain topics to avoid emotional distress or over-involvement. P17 explained \textit{``while certain questions seemed helpful, continuing to engage with them might have deepened my emotional engagement, but I wasn't quite ready at the moment. I thought I would come back at some point.''}}
Based on the findings from debriefing interviews, we present how participants interacted with the themes and questions in the pathways of processing their thoughts and emotions.

\subsubsection{Expanding Viewpoints through Themes} 

Participants generally found the LLM-generated themes were reflective and responsive to their previous input, with many noting the relevance and accuracy of the suggestions. They reported that they selected the themes by prioritizing clarity and personal relevance, aiming to focus on issues that they felt were actionable and meaningful. For instance, P18 preferred more \textit{``condensed''} words as themes, believing they would lead to more insightful AI-generated questions. Others, like P5 and P8, tended to select themes that aligned with their immediate concerns, allowing them to focus on issues they could actively address. At the same time, however, participants avoided themes that felt overly literal or much echoed their initial expression. For example, P2 preferred themes that \textit{``offered some interpretation or deeper insight,''} because she wanted to \textit{``explore the heart of the matter, rather than focusing on surface-level concerns.''} This balance between relevance, clarity, and depth helped participants feel more in control of their reflective process. 

Participants also emphasized that the suggestions of alternative wordings for the same theme allowed them to engage in their narrative from multiple angles, broadening their perspectives and deepening their reflections. This functionality encouraged participants to think about the same theme in different ways, as seen when P14 explored the theme of `retirement' from multiple viewpoints, such as ``\textit{anxiety about retirement},'' ``\textit{purpose and direction of life after retirement},'' and ``\textit{feeling unprepared for old age.}''
These varied expressions also provided flexibility in how participants responded to the themes, prompting different sets of questions depending on how a theme was framed, and also expanded their capacity for self-expression. As P9 noted, the system's varied prompts allowed him to \textit{``respond more creatively and diversely''} by offering new ways to articulate their thoughts.

Participants tended to move on to a new theme when they felt they had gained enough insight or a good understanding of a particular issue. For example, P13 shifted her focus toward solutions once she understood that her low self-esteem was the root of her challenge: ``\textit{I decided to think about how to address it, rather than keep going deeper into why my self-esteem was low.}''

\subsubsection{Organizing Angles and Attitudes through Questions}
Participants found that being presented with multiple questions helped them consider their challenges from various angles, exploring different aspects and depths of their situation. P7 noted how it allowed him to see \textit{``how different questions can expand one thought into multiple pathways,''} likening the process to a \textit{``mind-storming''} session where diverse questions spurred deeper thought. P11 added that seeing questions tied to his previous responses helped \textit{``take the reflection process step by step, encouraging [him] to explore aspects they may not have considered otherwise.''} Some participants viewed the act of deciding which question to engage with based on what seemed most effective at the moment as a reflective practice in and of itself. P5 mentioned, \textit{``If new challenges arise, I can think through them using these types of questions, such as considering why the problem occurred, what abilities I have to resolve it, and how I might approach the solution.''} 


The provision of multiple follow-up questions enabled participants to steer the reflection process, often balancing their cognitive and emotional discomfort with the question's significance regarding their personal concerns. Many chose questions that carried the most weight on their issues, even if those questions were mentally burdensome to answer. As P18 shared, she prioritized \textit{``the most pressing concern,''} recognizing that addressing these difficult topics would ultimately help her confront more difficult issues.

Participants also valued questions that challenged their usual thinking patterns, prompting reflections on aspects they had never considered before. P2 noted, \textit{``I usually set personal goals and feel frustrated when I don't meet them. The AI asked, ‘Why do you set such standards?’ which made me select it because it questioned something I had always taken for granted, like the fact that the Earth is round. It was an idea I had never thought about, and that’s why it stood out to me.''}
In contrast, some participants avoided more abstract or challenging questions, opting for those that \textit{``felt easier to answer based on concrete experiences (P5).''} To balance the cognitive load, several participants voluntarily alternated between engaging with challenging, thought-provoking questions and those they found more manageable. \deleted{For example, P5 noted, \textit{``It wasn’t easy to think about, but it [ExploreSelf] asked `what my strengths are in a competitive setting,' so I went ahead with [that question] because I figured it would be useful to learn about,''} expressing willingness to confront difficult questions if he believed the outcome would ultimately be beneficial to him. Similarly, P8 shared that after writing about difficult and negative experiences, she felt emotionally drained. When presented with multiple follow-up questions, she opted for one that could \textit{``shift their thinking in a new direction''}, selecting \textit{``How would your life change if your financial situation were stable?''} over other equally \textit{``relevant and effective questions''} such as, \textit{``What impact is your financial difficulty having on your life?''} or \textit{``What steps have you taken to resolve your financial challenges?''}. This choice allowed her to regulate her emotional state while still maintaining engagement without becoming overwhelmed by the difficult emotions.}

\subsubsection{Objectification and Reiteration through AI Summary}
Participants generally viewed the AI summary to bring clarity and objectivity to their often complex and unorganized thoughts by distilling the fragmented responses into an organized narrative. In turn, the AI summary seemed to foster iterative reflection; twelve out of 19 participants (63\%) revisited the Exploration page, explored more, and went back to the AI summary page to check the new summary reflecting additional responses (See dark blue bars in \autoref{fig:results:durations}). P12 remarked, \textit{``I liked the summary, but I was curious how it might turn out if I revisited it, so I kept checking. After seeing the summary again, I thought of things I had missed and wrote them down, then went back to see how the updated summary looked. This back-and-forth really helped me organize my thoughts.''} This iterative process, where participants reviewed and refined their reflections, enhanced their engagement and allowed for further self-exploration.

Some participants also reported that the AI summary helped them face uncomfortable or avoided thoughts. P18 recalled that she initially wrote, \textit{``Watching YouTube videos makes me more tired,''} and seeing the AI repeat it in the summary forced her to confront this feeling more directly. The AI’s neutral tone in the summary helped participants face these challenging thoughts without feeling overly judged, as P2 also described, \textit{``It’s something I already know, but the AI organizes my situation objectively. If a person did this, it might feel uncomfortable, but with AI, it doesn’t bother me.''}

\subsection{Supporting Self-Expression}
On the question panels, \sysname{} provided adaptive guidance, including keywords and comments. In this section, we cover how these elements of guidance helped participants' exploration and reflection process.

\subsubsection{Enriching and Rethinking Expressions through Keywords}
Participants could toggle on the keywords to reveal them, and seventeen out of 19 (89\%) participants saw keywords more than once during the exploration phase (see `Total \# of keywords' in \autoref{tab:usage-data}), but with varying levels of engagement with them. Some participants received keywords as a valuable tool for articulating their thoughts, while others either used it selectively or chose not to rely on them. 
Many participants used the keywords as prompts to help express their emotions and ideas more clearly. For instance, P7 mentioned while reflecting on emotions after a conflict with their children, he initially thought of simplistic terms like \textit{``anger''} or \textit{``sadness.''} However, the keywords suggested deeper and more nuanced emotions like \textit{``regret''} and \textit{``powerlessness,''} which broadened their emotional vocabulary and helped express himself more fully. Similarly, P14, emphasizing that her age is over 60, mentioned that \textit{``As I age, I struggle to think of the right words and expressions, but seeing the keywords made it easier to express myself.''}

Some participants used the keywords as a way to explore new perspectives or deepen their reflections. P8, for instance, frequently clicked \textit{See More Keywords} by mentioning, \textit{``the AI was smarter, showing me words I couldn't have thought of.''} The keywords would prompt participants to think beyond their initial responses and anchor their reflections. P13 also mentioned that keywords sparked associations with aspects of the matter that she had not thought of in the first place. 
At the same time, some participants reported that the keywords mitigated the mental burden while organizing negative thoughts for answering. P4, who had experienced trauma, found the keywords helpful in guiding him through some of his painful memories. He noted that the keywords made it less emotionally overwhelming to revisit these thoughts because \sysname{} offered possible expressions they could choose from, making the process feel more manageable.

\subsubsection{Iteratively Developing and Refining Answers through Comments}

Participants were exposed to one comment per question by default, and fifteen out of 19 (79\%) requested additional comments more than once during the exploration phase (see `Total \# of comment requests' in \autoref{tab:usage-data}), with varying number of requests per participant. Initial impressions of the comments seemed to shape how participants continued to use or avoid the feature. For instance, P17 mentioned that while they initially had low expectations, receiving an encouraging comment led them to revisit the feature more often, curious about what other positive feedback they might receive. On the other hand, P9, who has requested comment only once, noted that after receiving a comment that felt irrelevant to their context, they decided not to use the feature again.

Many participants appreciated the responsiveness of the comments, which provided a sense of dynamic interaction. For instance, P19 noted that the comments felt like having a \textit{``running mate,''} offering feedback that helped him stay engaged. P13 similarly expressed curiosity about how the comments would respond, motivating her to continue interacting. P11 described how the comments helped refine his thoughts, iteratively guiding him from a single word, \textit{``financial stability''}, to a more developed sentence. The comment prompted him to \textit{``think about what you ultimately want from financial stability,''} which led to a detailed response about \textit{``owning property, a car, and a happy family.''}

\revised{Comments generally provided participants with serendipitous insights through a more detailed message, and participants thought a comment was particularly insightful when it acknowledged their feelings or thoughts with clear rationales: ``\textit{[...] The fact that it acknowledged my feelings as natural was really meaningful to me (P7).}''
Although the encouraging tone of comments often engaged participants in the exploration, not all experiences were positive. For example, P16 felt that some comments were overly upbeat and didn’t match the gravity of her situation, causing a sense of emotional disconnection.}

\deletedsubsection{Perceived Utility of ExploreSelf}
%
\deleted{To gauge the utility of ExploreSelf beyond the study, we asked participants about their willingness to try ExploreSelf in the future. All but one participant (18 out of 19; 95\%) said that they would like to use ExploreSelf after the study to process the challenges that would arise in the future. One participant chose not to use the tool because he had expected an AI to offer actionable solutions for his challenges. Others envisioned revisiting ExploreSelf when they have recurring struggles \textit{``like a tangled ball of yarn''} (P9).}

\section{Discussion}
In this section, \revised{we discuss lessons learned from the design and implementation of \sysname{} as well as from the user study. We also reflect on the implications for better supporting LLM-guided self-reflection in a long-term and inclusive setting.}

\begin{revisedenv}
\subsection{Supporting User-driven Exploration for Mental Well-being}

\sysname{} facilitated user control over their reflective journey by offering multiple options and layers of adaptive support that encouraged users to navigate their thoughts and emotions.
This scaffolded structure allowed participants to tailor their exploration pathways, balancing their cognitive and emotional burdens while facing various aspects of their personal challenges.
For example, participants chose to engage more deeply with a theme through Socratic questions or pivoted to new themes when they felt ready, demonstrating varied patterns of exploration (see \autoref{fig:results:flow_participants}).
Consequently, participants showed a significant increase in perceived agency after engaging with \sysname{}, underscoring its potential to facilitate for users to manage their reflective journey.

Our main design goal with \sysname{} was to prioritize user agency without directing content, while providing guidance. This goal resulted in the emphasis on \textit{exploration}, leading to design choices that are different from prior LLM-assisted reflection tools, such as DiaryMate~\cite{kim2024diarymate} and MindfulDiary~\cite{kim2024mindfuldiary}. For example, DiaryMate employed the concept of writing assistant~\cite{lee2024designspacein2writing} and used LLMs to generate diary sentences given user inputs like keywords and the model temperature~\cite{kim2024diarymate}. While this approach helped participants complete well-written journal entries, the quality and content of LLM generations would largely influence their reflection and even sometimes interrupt their own stream of thought~\cite{kim2024diarymate}. 
MindfulDiary incorporated a conversational agent, loosely mimicking a counseling interview with users ~\cite{kim2024mindfuldiary}. Although the conversational approach effectively demonstrated care and empathy, the LLM-driven agent mostly held the initiative of conversation and the turn-taking interaction focused the user's attention on a single question presented at each turn~\cite{kim2024mindfuldiary}.
\sysname{} differs from previous work in that, unlike DiaryMate, it offers users choices about what to explore (\ie, themes and questions) rather than generating narratives on their behalf. Additionally, users could determine their level of engagement with each topic through an interface designed to minimize single-path conversational affordances, as in MindfulDiary. This approach encouraged users to `contemplate' their exploration, enabling user-directed control over the process.

In this work, we explored a novel design approach to invite user agency throughout a self-paced exploration for mental well-being.
\sysname{} targeted lay individuals who may not require clinical diagnoses but any of those who would willingly seek assistance in navigating personal challenges. In this context, its approach follows the principles of psychotherapy, particularly its emphasis on client autonomy~\cite{Scheel2011ClientCommonFactors}. Guided by self-determination theory~\cite{deci2013intrinsic}, it is widely agreed that supporting clients in autonomously exploring and sustaining a process of change is crucial for a successful outcome~\cite{ryan2008self, Scheel2011ClientCommonFactors, patall2008effects}. Similarly, ExploreSelf provided an environment where users can process their challenges at their own pace and on their own terms, which mirrors the role of therapist---fostering a space where individuals feel empowered to explore their emotions, identify goals, and develop strategies for personal growth~\cite{Ryan2011Motivation}---illustrating the potential of designing adaptive guidance with LLMs.

\end{revisedenv}

\subsection{Design Considerations for Long-term, Multi-session Interaction}

At an early stage of investigating the feasibility of leveraging LLM-guided exploration of personal challenges, our work was based on a single-session lab study. That being said, \revised{in the debriefing,} all participants except one \revised{expressed willingness to engage with} \sysname{} continuously in the future, particularly \revised{in circumstances where they would need} to process complex thoughts or emotions. For example, P14 imagined the system could help users recognize recurring themes, providing deeper insights into long-standing thought patterns and behaviors. Their feedback underscores \sysname{}'s potential to \revised{serve as a channel for people to engage with regularly to process their} personal challenges. 

\revised{Supporting long-term engagement with \sysname{}---revisiting \sysname{} in multiple sessions over time---requires a new set of design considerations. The current version of \sysname{} expands the branches of reflection rooted in the initial narrative, which the user provides in a prelude to the exploration. Intuitively treating this unit as a session, users may create multiple sessions or append more themes to a previously-explored session. Notably, the sessions of the same user may be interconnected, sharing common aspects of and sometimes influencing one another. For example, a series of personal challenges can lead to the discovery of an overarching theme. Also to ensure accuracy and reliability,} the system should incorporate sort of \textit{long-term memory} (\cf{}, \cite{jo2024carecall_ltm, bae2022keepmeupdated}), where it leverages the information derived from past sessions for the current session. For example, the system may memorize the interpersonal relationships of the user and consult it when generating themes and questions accordingly. However, simply stitching the entire interaction history for model input is neither scalable nor reliable~\cite{bae2022keepmeupdated}\revised{, and could lead to inaccurate or irrelevant summaries and guidance}. This calls for future research in designing how the system constructs and synthesizes the user knowledge that is noteworthy. 

Also, prolonged user engagement with the system will also increase the volume of interaction history to be presented. Therefore, interaction techniques such as semantic zooming~\cite{Boulos2003semanticzoom} and the multi-level summarization approach would be necessary. The long-term data will also pose opportunities to present other supplementary data, such as recurring questions or themes that the user keeps avoiding exploring: P13 imagined, ``\textit{I noticed that I have avoided the questions on a specific topic. Seeing this allowed me to better understand myself about what I don't want to face.}'' Supporting such `byproduct' reflections calls for further investigation from the perspectives of the personal informatics systems~\cite{Li2011SelfReflection, Choe2017SelfReflection, Kim2021Data@Hand}.

\revised{To comprehensively support people in addressing personal challenges, future versions of \sysname{} should go beyond the exploration phase. Our user study results indicate that about 30 minutes of exploration fostered a certain level of sensemaking about the self and the subject matter. According to the transtheoretical model~\cite{Prochaska1997transtheoretical}, the exploration mostly supports individuals to reach the \textit{contemplation} stage, where they recognize what to change to address their challenge. However, different types of guidance and interaction may be more effective for those who have progressed to the \textit{preparation} stage, where they began taking small steps towards change.}

\begin{revisedenv}
\subsection{Role and Stance of AI in Guiding Reflection}
Although we minimized conversational affordances from the interface, \sysname{} still conveyed messages in human language with a tone through comments. Within the current scope of investigation, we followed a broadly applicable guideline and instructed the model to generate empathetic and supportive comments. However, the positive tone of comments was sometimes perceived as inconsiderate of the gravity of the participants' concerns. This underscores the need for a more nuanced design of the AI's role and stance embedded in its messages. While \sysname{} is not intended to mimic psychotherapy, the therapists from our formative interviews highlighted that balancing both empathy and confrontation is often a helpful therapeutic tactic, pointing out that they do not always show agreement and support to their clients in leading them to positive change. Notably, counselors sometimes \textit{confront} their clients with discrepancies in their perception of reality ~\cite{anderson1968confrontation}, which may lead to resistance~\cite{resnicow2012motivational}.

Applied to \sysname{}, the AI may take a stance that properly facilitates reflection on the fly that is aligned with these principles.  To enable users to engage with confronting guidance when they are mentally ready, the system may provide a set of different stances that users can select as needed. However, LLMs, particularly those fine-tuned with human alignment techniques like RLHF~\cite{ouyang2022instructgpt}, typically exhibit supportive, empathetic, and cooperative behaviors~\cite{wei2023leveraging, lee2024empathetic, zhu2024personalityalignment} and this tendency is difficult to override by prompting alone~\cite{li2024big5chat, zheng2024helpfulassistant}. To demonstrate a confronting stance, the system may leverage customized LLMs fine-tuned with annotated transcripts from psychotherapy sessions such as Motivational Interviewing~\cite{resnicow2012motivational}.
Here, users' interaction logs would also play a key role in assessing their resistance to specific themes. For instance, P12 and P14 noted that reviewing unselected questions might uncover the aspects of the challenge that he often avoids contemplating. A future version of \sysname{} can analyze the patterns in these \textit{roads not taken} and determine the level of confrontational guidance provided to the user.
%


\subsection{Supporting Severe and Sensitive Challenges}
Observing participants exploring diverse personal challenges of varying severity, we identified implications for supporting individuals with severe challenges, including those related to mental illness, such as depression, traumas, or other emotional distress. 
For individuals experiencing severe apathy, commonly seen in conditions like depression, self-directed exploration may be overwhelming~\cite{smith2022lower, kim2024mindfuldiary}. This echoes the feedback from experts in our formative interviews, who noted that individuals with severe emotional numbness may struggle to engage meaningfully with open-ended reflective tools. In such cases, more proactive guidance can be more effective than a high degree of choice. 

Similarly, reflecting on one's trauma necessitates careful guidance due to its emotional intensity involved~\cite{perlman1998therapist, miller2001creating}. To keep the exploration within the emotionally safe boundary, \sysname{} may receive traumatic keywords or precautions from users in advance to consider them when handling the emotionally charged material. Although \sysname{} allows users to decide when to stop drilling into certain themes, individuals exploring a traumatic challenge may become overly involved in the exploration. Future work can investigate whether and how the system should intervene during exploration. However, simply terminating a session, particularly when users lack alternative resources, raises a lot of ethical concerns about potentially leaving individuals without support in critical moments. Instead, it is essential to develop guardrails that facilitate navigation for vulnerable users while ensuring their safety. This is especially important given that many individuals increasingly turn to accessible LLM interfaces for coping with mental health challenges in the absence of other available options~\cite{song2024typing}.


Cultural awareness is crucial for addressing deeply complex and sensitive challenges, as cultural consideration is an integral aspect of mental health support~\cite{pendse2019mental, balcombe2022human, christopher1999situating, lent2004toward}. Considering the technical reliability and availability, we selected OpenAI GPT---an LLM trained with datasets dominated by English and Western cultural references---as the underlying model for our generative pipelines in Korean. Although we did not receive participants' feedback that the AI generations lacked sensitivity to Korean cultural norms, the system may still reflect certain Western cultural norms and biases~\cite{pawar2024survey, tao2024cultural} when handling challenges specific to non-Western contexts. This limitation can be particularly problematic when engaging with marginalized communities, for whom issues such as racism or sexism are not mere distortions but ongoing harms~\cite{haslanger2017racism, mcgowan2019just}. Employing LLMs aligned with the language and cultural context of the target population (\eg{}, HyperCLOVA~\cite{Kim2021HyperCLOVA} in Korean) could better reflect culturally relevant references~\cite{seo2024chacha, naous2024culturalbiasLLM}. However, the disparity in the model performance of LLMs across different languages and regions remains an open challenge.

\end{revisedenv}

\subsection{Limitations and Future Work}
In this section, we discuss the limitations of our study that may impact the generalizability of the findings.
First, our study participants might not have fully disclosed their innermost challenges, given it is a research study setting. We tried to address this issue by: (1) minimizing the invasiveness of observation by creating breakout rooms where participants can have their privacy in exploration; and (2) collecting and observing only the interaction logs, not the actual problem narratives to respect their privacy. To further ensure their safety, we limited the time of exploration so that the study would not get too emotionally charged for our participants, and prepared safety protocols in case of emergency. 
We also acknowledge that collecting data from more ecologically valid settings (\eg{}, field deployment) may yield different distributions of topics and exploration patterns. 
Furthermore, although we aimed to recruit participants from diverse backgrounds, all participants are residents in South Korea. Populations with different socio-cultural contexts and AI literacy may engage in different types of personal challenges using \sysname{}.

\deleted{Considering the technical reliability and its availability, we selected OpenAI GPT---an LLM trained with datasets dominated by English and Western cultural references---as an underlying LLM for our generative pipelines in Korean. Although we did not encounter participants' feedback that the AI generations did not correspond with Korean cultural sensitivity, using LLMs aligned with the language and cultural context of the study population (\eg{}, HyperCLOVA in Korean) may potentially improve the quality of generation.}
\section{Conclusion}
We presented \sysname{}, an LLM-driven application that supports processing personal challenges by exploring related themes and questions. \sysname{} enables people to take the initiative of exploration, thereby maintaining their agency and proactively selecting what aspects they want to explore. To examine how people use \sysname{} to explore and reflect on their own challenges, we conducted an exploratory user study with 19 individuals. Participants successfully adopted the concept of proactive exploration interactions that \sysname{} offered, showing various patterns. Our study revealed that providing multiple options with adaptive guidance enhanced participants' perceived agency, and the large language model played a key role by providing wordings that acknowledged the participants' language in detail.
\revised{We highlighted several areas for future research, including designing LLM-guided self-reflection systems that balance autonomy and intervention, technical and design considerations for long-term reflective practices, and incorporating cultural and emotional sensitivity to address diverse user needs and challenges.}
In summary, our work contributes to the growing body of LLM-driven mental well-being support, demonstrating the feasibility and usefulness of flexible user interface inspired by reflective writing. 

\begin{acks}
We thank our participants in both formative interviews and the user study for their time and efforts. We are also grateful to Eunjoo Kim at Yonsei University and Kyungah Lee at Dodakim Child Development Center for helping us recruit professionals for our formative study. We also thank Dasom Choi for her feedback on the early version of our draft.
This work was supported through a research internship at NAVER AI Lab of NAVER Cloud.
\end{acks}

\bibliographystyle{ACM-Reference-Format}
\bibliography{bibliography}

\newpage
\onecolumn
\appendix

\section{Formative Study Material}
\subsection{Example Narrative}~\label{app:formative-example}

\textit{Ever since middle school, I’ve always seen myself as someone who thrives in the shadows, away from the glaring spotlight of public attention. I preferred to keep my head down, diligently working in quiet solitude, and my social interactions were always subtle, never the center of attention. But recently, my usual tranquility at work was disrupted when I was unexpectedly chosen to lead a major project. This wasn’t just any project; it involved a division-wide presentation, thrusting me into the very spotlight I had always avoided. The thought of standing in front of all those people, having to lead and direct, filled me with an intense dread that was completely new to me.}

\section{Summarized Task Instructions for Each Generation Pipeline}

\subsection{Theme Generation}~\label{app:theme-gen}
\ttfamily
\begin{itemize}
    \item Role: You are a therapeutic assistant designed to foster deep self-reflection and promote personal growth in clients. Your approach is empathetic, client-centered, and rooted in the principles of therapeutic inquiry.
    \item Input type and format: 
    
    <initial\_information/>: Client’s initial brief introductory of difficulty narrative, and the client’s background.

    <previous\_session\_log>: Logs of sessions before the current session. DO NOT overlap with the previously selected themes.
    \item Output:
    \begin{itemize}
        \item main\_theme: Each theme from the user's initial narrative and previous log. Align closely with the user's language, expressions.
        \item expressions: An array of diverse different expressions of the main\_theme. When appropriate, introduce metaphoric expressions or nuanced language that might provide additional therapeutic value, but always anchor these in the client’s original phrasing and emotional context.
        \item quote: Most relevant part of the user's log to the theme
    \end{itemize}

\end{itemize}
\rmfamily

\subsection{Question Generation}~\label{app:question-gen}
\ttfamily
\begin{itemize}
    \item Role: You are a therapeutic assistant specializing in generating socratic questions to facilitate self-reflection and personal growth in clients. Per each session within the system, the client brings up a Theme in one's narrative that one would like to navigate about.
  \item Task: Given a client's personal narrative and context, your task is to generate list of Socratic questions and intention of the question
  \item Input type and format:\\
  <initial\_information/>: Client's initial brief introductory of difficulty, and the client's background.\\
  <previous\_session\_log>: Logs of sessions before the current session. \\
  <theme\_of\_session/>: Theme of the current session. 
\end{itemize}
\rmfamily

\subsection{Keyword Generation}~\label{app:keyword-gen}
\ttfamily
\begin{itemize}
    \item Role: You are a therapeutic assistant specializing in supporting users to response to Socratic questions, facilitating self-reflection and personal growth. 
  \item Task: The user is given a Socratic question to think about. However, it's not cognitively easy to answer those questions.
  Therefore, your task is to provide 'keywords' or 'short phrases' that might be useful for user to answer the given question. These keywords might act as 1) cognitive scaffolding 2) activate what might have been blind spot of the user 3) what might be relevant to users context 4) and so on.
  \item Input type and format:\\
  <initial\_information/>: Client's initial brief introductory of difficulty, and the client's background.\\
  <previous\_session\_log>: Logs of sessions before the current session.\\
  <question/>: Question that the user is trying to response now.  
\end{itemize}
\rmfamily

\subsection{Comments Generation}~\label{app:comment-gen}
\ttfamily
\begin{itemize}
    \item Role: You are a therapeutic assistant specializing in supporting users to work on personalized self-help workbook.
  \item Task: While the user is responding to the Socratic question, your task is to provide useful comments based on the context of user's response.
  For example, if the user hasn't started answering the question or is in very early phase, you might a) provide tips on how to approach the question.
  You might also provide b) encouraging, affirmation, or supportive feedback in a user-tailored manner. It also could be the c) sub-question in answering the given question, or some d) insightful comment, or e) any other type of comment that might support user in responding to the question.
  \item Input type and format:\\
  <initial\_information/>: Client's initial brief introductory of difficulty, and the client's background.\\
  <previous\_session\_log>: Previous Logs of the session before the current session.\\
  <question/>: Question that the user is trying to think about now. \\
  <current\_response>: Current response status of the user
\end{itemize}
\rmfamily

\subsection{Summary Generation}~\label{app:summary-gen}
\ttfamily
\begin{itemize}
    \item Role: You are a therapeutic assistant helping the user reflect on their session and progress.
  \item Task: Summarize the user’s experiences and insights from their Q\&A log into a coherent and concise narrative. Focus on the essence of their reflections without overemphasizing any one aspect.
  \item Guidelines: 
  \begin{itemize}
      \item Capture the key points and overall sentiment without unnecessary detail.
      \item Use the user’s own language and expressions where appropriate.
      \item Keep the summary realistic, proportional to the content of the user’s log, and grounded to user's input.
      \item Feel free to draw on the following as needed
      \begin{itemize}
          \item Major themes or emotions
          \item Notable progress or changes in perspective
          \item Encouragement to continue reflecting and growing
          \item Recognition of the user’s strengths and resources
      \end{itemize}
  \end{itemize}

\end{itemize}
\rmfamily

\section{Statements of Pathway Subscale of Adult State Hope Scale}~\label{app:pathway-scale}

\noindent{}The validated Korean version~\cite{choi2008validation} of the Pathway subscale of the Adult State Hope Scale~\cite{snyder1996development} consists of the following four questions:

\begin{enumerate}
    \item If I should find myself in a jam, I could think of many ways to get out of it.
    \item There are lots of ways around any problem that I am facing now.
    \item  I can think of many ways to reach my current goals.
    \item I know that even when others despair, I can find a way to solve my problems.
\end{enumerate}

\noindent{}Each question is rated on an 8-point rating scale, with scores and labels as in the following:
\begin{enumerate}
    \item Definitely False
    \item Mostly False
    \item Somewhat False
    \item Slightly False
    \item Slightly True
    \item Somewhat True
    \item Mostly True
    \item Definitely True
\end{enumerate}

\end{document}